\documentclass[preprint, nofootinbib, superscriptaddress]{revtex4}
\usepackage{graphicx,subfig,hyperref}
\usepackage[normalem]{ulem}
\usepackage{xcolor}
\usepackage{amssymb}

\begin{document}
\title{Non-radial oscillations of hadronic neutron stars, quark stars, and hybrid stars : Calculation of $f$, $p$, and $g$ mode frequencies}
\author{Atanu Guha}
\affiliation{Department of Physics, Chungnam National University, Daejeon 34134, Korea}
\author{Debashree Sen}
\email{debashreesen88@gmail.com}
\affiliation{Center for Extreme Nuclear Matters, Korea University, Seoul 02841, Korea}
\author{Chang Ho Hyun}
\affiliation{Department of Physics Education, Daegu University, Gyeongsan 38453, Korea}

\date{\today}
\begin{abstract}
The composition and equation of state (EoS) of dense matter relevant to compact stars are quite inconclusive. However, certain observational constraints on the structural properties of compact stars help us constrain the EoS to a fair extent. Moreover, gravitational asteroseismology gives a notion of the composition and EoS of compact stars. The next generation gravitational wave (GW) detectors are likely to detect several oscillation mode frequencies of the GWs. In this work we compute the fundamental ($f$) and the first pressure ($p_1$) mode frequencies ($f_f$ and $f_{p1}$, respectively) with different compositions viz., hadronic, quark, and hybrid star (HS) matter. For HSs, we also study the gravity ($g$) mode frequency ($f_g$). For each phase we also study the correlation between the oscillation frequencies of 1.4 $M_{\odot}$ and 2.01 $M_{\odot}$ compact stars with other different properties. We find that various possible composition of compact stars substantially affects the oscillation frequencies. However, the mass-scaled angular $f$ mode frequency ($\omega_f M$) varies universally with compactness ($C$) for all hadronic, quark and hybrid stars. The $f$ mode frequency ($f_{f_{1.4}}$) of the canonical 1.4 $M_{\odot}$ compact star, obtained with different composition, is quite correlated with the canonical radius ($R_{1.4}$) and tidal deformability ($\Lambda_{1.4}$) while $f_{p_{1.4}}$ is well correlated with slope parameter of the symmetry energy. We also show that $f_{g_{1.4}}$ of the HSs varies almost linearly with $\Lambda_{1.4}$. Should $g$ modes be detected, they could not only support the existence of HSs, but $f_g$ could be useful to understand the strength of quark repulsion in HSs.
\end{abstract}

\maketitle

\section{Introduction}
\label{Intro}

Several extreme conditions of density ($\rho$ = 5 - 10 times nuclear matter density), compactness ($C$ = $M/R$), and etc. related to the neutron star (NS) environment make them one of the most fascinating celestial objects to study. However, till date a lot of uncertainties still pertain to their equation of state (EoS), which is developed on the basis of theoretical modeling of NS matter. These uncertainties are generally attributed to the fact that the composition, properties and interaction of matter at such high density are largely inconclusive. Over several decades theoretical speculations suggest that NS cores are dense enough to sustain the existence of exotic matter and even signatures of hadron-quark phase transition thereby forming hybrid stars (HSs). In order to achieve phase transition, mechanisms like Gibbs and Maxwell constructions \cite{Maruyama:2007ss,Orsaria:2019ftf}, hadron-quark crossover \cite{Qin:2023zrf,Constantinou:2021hba,Sotani:2023zkk} and constant speed of sound parameterization \cite{Sun:2023glq,Laskos-Patkos:2023cts} etc. are often adopted in literature. The Bodmer-Witten conjecture suggests that strange quark matter (SQM), being more stable than pure nucleonic matter, is the true ground state of the hadronic matter \cite{Farhi:1984qu,Torres:2012xv,Olinto:1986je}. This implies that the entire compact star may be composed of SQM and thereby strange quark stars (SQSs) may also exist. At very high density, QM can also lead to the formation of color-flavor-locked (CFL) QM \cite{Rajagopal:2000ff,Alford:2002rj,Lugones:2002va,VasquezFlores:2017uor}. Consequently, on the basis of the composition and interactions, we obtain the EoS of hadronic neutron stars (HNSs), SQSs, and HSs. Each class of EoS has its own uncertainty revolving around the consideration of composition and interactions. With the recent developments in the observational sector, the EoS has been constrained to a certain extent. The most important of such observational constraints include the detection of gravitational wave GW170817 by the LIGO/Vigro Collaboration \cite{LIGOScientific:2018cki} and the simultaneous measurement of mass and radius of HESS J1731-347 \cite{Doroshenko:2022} and those of PSR J0030+0451 \cite{Riley:2019yda,Miller:2019cac} and PSR J0740+6620 \cite{Fonseca:2021wxt,Miller:2021qha,Riley:2021pdl} by NICER experiment.

With the recent advancements of the upcoming gravitational wave (GW) detectors such as the LIGO O4 run, the Einstein Telescope \cite{Punturo:2010zz,Hild:2010id}, and the Cosmic Explorer \cite{Reitze:2019iox,Evans:2023euw}, it can be expected that the oscillation mode frequencies of the GWs can be detected. This will provide deeper insight into the interior of the compact stars. Such oscillation spectra of GWs consist of different modes viz., fundamental ($f$), pressure ($p$), rotational ($r$), space-time ($w$) and gravity ($g$) modes \cite{Kokkotas:1999bd}. The restoring force that re-establishes the equilibrium is different for each mode. In case of both the $f$ and $p$ modes, the equilibrium is restored by the fluid pressure of the star, and for the $g$ mode, gravity acts as the restoring force. The efficiency of the upcoming detectors is expected to help the branch of GW asteroseismology, especially in terms of the most prominent $f$ mode frequency \cite{Abac:2025saz}. This is because compared to the other modes, $f$ mode has the strongest tidal coupling and below resonance the behavior of $f$ mode is the primary contributor to the tidal deformability. Therefore, the detection and measurement of the $f$ mode frequency is the most plausible in the upcoming detectors. The other modes, especially the $g$ mode, are comparatively difficult to detect because their tidal couplings are much weaker. Considering the advances in the upcoming detectors, it is important to compute the oscillation mode frequencies with the different EoS. Several studies have been conducted on these different oscillation modes \cite{Sotani:2010mx,Flores:2013yqa,Andersson:2019mxp,Zhao:2022tcw,Kumar:2024jky,Ranea-Sandoval:2018bgu,VasquezFlores:2019eht,Pradhan:2022vdf,Jyothilakshmi:2024zqn,Das:2021dru,Hong:2023udv,Shirke:2024ymc,Jaiswal:2020wzu,Constantinou:2021hba,Orsaria:2019ftf,Kumar:2021hzo,Zheng:2023oba,Sen:2024yim,Wen:2019ouw,Sotani:2021nlx,Jyothilakshmi:2024xtl,Thakur:2024ijp} based on different compositions of the star. Several works dedicated to this topic suggest strong correlation between the $f_f$ and compactness \cite{Andersson:1997rn}, moment of inertia \cite{Lau:2009bu} and static tidal polarizability \cite{Chan:2014kua,Sotani:2021kiw} for NSs, QSs and HSs \cite{Zhao:2022tcw}. This leads to the emergence of the universal relations between such properties of compact stars with $f_f$. $p$ mode frequency ($f_{p1}$), on the other hand, are weakly correlated with such properties of NSs \cite{Andersson:1997rn}. Moreover, they are largely affected by the EoS of the crust, unlike the $f_f$, especially at low density \cite{VasquezFlores:2017tkp,Pradhan:2020amo,Kunjipurayil:2022zah}. At zero temperature ($T$=0) the $g$ mode and its corresponding frequency $f_g$ arise due to the discontinuity in density of the NS, which may be attributed to the phenomena like first order phase transition. Thus it can be described by HS EoS \cite{Ranea-Sandoval:2019miz,Flores:2013yqa,Sotani:2010mx,Ranea-Sandoval:2018bgu,Tonetto:2020bie,Wei:2018tts,Rodriguez:2020fhf,Kumar:2023rut,Zhao:2022toc,Tran:2022dva,Zhang:2023zth}. Universal relations are found to exist between the mass-scaled $g_1$ mode frequency and the radius-scaled $f$ mode frequency; and also between the compactness-scaled $g_1$ mode frequency and tidal deformability \cite{Kuan:2022bhu}. For an isolated, self-bound star, the $g$ mode has weakest tidal coupling and $g$ mode emission in the GWs is much weaker than the preceding modes. Therefore, it becomes quite challenging to detect $g$ mode in near future. However, in a special case of binary merger, that involves at least one HS, the tidal energy of the resonant $g$ modes can be twice or thrice the same for normal NS. This can lead to considerable excitation of the $g$ mode that may aid the detection of $f_g$ in future \cite{Jaikumar:2021jbw}. 

In the present work, we study the effects of the different composition of compact star matter on the non-radial oscillation frequencies of the $f$ and $p$ modes of the HNSs and SQSs with different models. For HSs we also calculate the $g$ mode frequency $f_g$. The first development of the theory of non-radial oscillation of NS in the framework of general relativity (GR) is done in \cite{1967ApJ...149..591T} while \cite{Lindblom:1983ps} first integrated the numerical solution of the NS oscillation. Later, the metric perturbations were neglected with the help of Cowling approximations \cite{Cowling:1941nqk} in order to obtain a simplified theory \cite{Sotani:2010mx}. The calculation of the oscillation frequencies in full GR conditions that include the spacetime oscillation also shows the emergence of another mode viz., the $w$ mode. Therefore, many recent works have calculated the oscillation frequencies in the framework of the total GR theory \cite{Shirke:2024ymc,Mondal:2023wwo,Zhao:2022tcw,Kunjipurayil:2022zah,Pradhan:2022vdf,Flores:2024hts,Tsui:2004qd}. The results of the calculation of the oscillation frequencies using Cowling approximation and total GR calculations can differ up to 30\% for the $f_f$, and $\sim$15\% for $f_{p1}$ \cite{Kunjipurayil:2022zah}. However, the overestimation with Cowling approximation can be less than 20\% from those calculated by considering the full
GR treatment \cite{Pradhan:2022vdf,Yoshida:1997bf}. Also, \cite{Tonetto:2020bie,Zhao:2022toc} showed that the Cowling approximation is reasonably good for the calculation of $f_g$. Overall, the gross qualitative results remain unaffected with or without Cowling approximation. Therefore, it is still widely adopted in the recent literature \cite{Kumar:2023rut,Jyothilakshmi:2024zqn,Arbanil:2023yil,Thakur:2024ejl,Thapa:2023grg}. In the present work we also consider the Cowling approximation in order to provide the estimates of different oscillation frequencies. For the purpose of exploring the uncertainty due to the EoS, we consider the KIDS (Korea-IBS-Daegu-SKKU) functional \cite{Gil:2020wqs,Gil:2020wct,Gil:2021ols}, and the relativistic mean field (RMF) models GM1 \cite{Glendenning:1991es} and DD2 \cite{Typel:2009sy} hadronic models. For SQSs we consider the vector Bag (v-Bag) model \cite{Torres:2012xv,Lopes:2021jpm}, the Nambu-Jona-Lasino (NJL) model with vector interaction \cite{Lopes:2020rqn,Masuda:2012ed}, and the CFL model based on the MIT Bag model \cite{Lugones:2002va}. For HSs, we follow Maxwell construction \cite{Kumar:2023lhv,Qin:2023zrf,Laskos-Patkos:2023tlr} with the KIDS model to describe the hadronic phase and the v-Bag model for the quark phase. We intend to study how the oscillation properties vary for the three different types of stars viz. HNSs, SQSs and HSs based on the composition. In this context, we mention  that the KIDS model is not considered before in literature to study the oscillation properties of HNSs and HSs.

We organize this paper in the following manner. In the next Sec. \ref{Sec:Models} we discuss the salient features of the different models considered in the present work on the basis of the different composition. Sec. \ref{Sec:phase transition} is devoted to the description of hadron-quark phase transition. We also provide the essentials of the mechanism to obtain the structural properties and different modes of non-radial oscillation of the compact star in Sec. \ref{Sec:Structure}. We display our results with corresponding discussion in Sec. \ref{Sec:Results}. We finally conclude in Sec. \ref{Sec:Conclusion}.


\section{Models for hadronic and quark matters}
\label{Sec:Models}

We consider the KIDS functional \cite{Gil:2020wqs,Gil:2020wct,Gil:2021ols} as the hadronic model to compute the 
structural properties and the oscillations frequencies of HNSs. We compare the results obtained with the KIDS model, with those of the two RMF models viz., the GM1 \cite{Glendenning:1991es} and the DD2 \cite{Typel:2009sy} models. 

The KIDS models are selected to have the ranges of the symmetry energy parameters
that are consistent with both nuclear data and astronomical observation of the neutron stars.
Symmetry energy parameters are defined by
\begin{equation}
S(\rho) = J + L x + \frac{1}{2} K_{\rm sym} x^2 + \cdots
\end{equation}
where $S$ denotes the symmetry energy, $x = (\rho - \rho_0)/3 \rho_0$, $\rho$ is the matter density and $\rho_0$ is the nuclear saturation density. The parameter most relevant to the NS properties is the slope parameter $L$, and the range obtained from the KIDS model is 47--66 MeV. This is consistent with empirical range 40--76 MeV \cite{dutra2012}, but the result of PREX-2 indicates that $L$ could be larger 80 MeV \cite{prex2}. The possibility of $L$ above the empirical range is accounted for by
considering two well-known RMF models viz., GM1 and DD2. Among them, GM1 has density-independent and DD2 has density-dependent nucleon-meson couplings. The two models are selected to have both similarity and difference with the KIDS models. In the DD2 model, the value of $L$ is within the empirical range, so it is similar to the KIDS model. However, the sign of $K_{\rm sym}$ is negative in the KIDS model, but positive in the DD2 model. Since $K_{\rm sym}$ also plays a critical role in the EoS of neutron star matter, the DD2 model can produce EoS stiffer than the KIDS model. Another RMF model GM1 has $L=94$ MeV, so one can easily expect that the model will give EoS much stiffer than the KIDS models. Thus, considering the four KIDS and two RMF models, the variation of $L$ ranges from 47 MeV (KIDS-D) to 94 MeV (GM1). The result, as will be shown soon, turns out as expected, so we can explore the impact of the stiffness of EoS to the oscillation mode.

To obtain the structural properties and the oscillations frequencies of the QSs we choose to work with different quark matter models viz., the vector Bag (v-Bag) model \cite{Torres:2012xv,Lopes:2021jpm}, the Nambu-Jona-Lasino (NJL) model with vector interaction \cite{Lopes:2020rqn,Masuda:2012ed}, and the color-flavor-locked (CFL) model based on the MIT Bag model \cite{Lugones:2002va}. 

All three quark models are quite well-known and have both similarities and differences among themselves. The unpaired quark models viz. the v-Bag and the NJL model have similarities that both the models consider the effect of vector interaction between quarks. However, the treatments of the two models are very different. The v-Bag model is based on the notion of confinement of the quarks with hypothetical `bag' while the NJL model considers the four-Fermi contact type of interaction. The CFL phase is also important at high density, which also introduces the role of the pairing gap. Therefore, the choice of the hadronic and quark models is based on different treatments. Therefore, we consider a variety of models in the context of treatment and properties to study their effects on the structural and oscillation properties. The salient features of each of the hadronic and quark models, considered in the present work, are briefly addressed below.

\subsection{Hadronic models}

\subsubsection{KIDS functional}
\label{Subsubsec:KIDS}

KIDS formalism starts from the density functional theory which states that the energy
of many-electron systems can be obtained as a function of the electron density from the first-principle calculation by using QED.
However, it is a difficult task to apply the first-principle calculation to the bound nucleon systems because of highly non-perturbative
nature of QCD at the energy scales of the bound nucleons. Therefore, it is hard to determine the functional form of the many-nucleon systems directly from QCD. Considering that the scale of momentum in nuclear medium is the Fermi momentum $k_{\rm F}$ and that the interactions are in intermediate and short ranges, one can apply the low-energy effective field theory to expand the energy per particle in nuclear medium in the power of $k_{\rm F}/m_\rho$ where $m_\rho$ is the rho-meson mass. Since $k_{\rm F}$ is proportional to $\rho^{1/3}$ in the cold multi-Fermion system where $\rho$ is the matter density, the energy per particle in homogeneous nuclear matter can be expanded as \cite{Papakonstantinou:2016zpe}
\begin{eqnarray}
{\cal E}(\rho,\, \delta) = {\cal T}(\rho,\, \delta) + \sum_{i=0}^2 \alpha_i \rho^{1+i/3} + \delta^2 \sum_{i=0}^3 \beta_i \rho^{1+i/3}.
\label{eq:kids}
\end{eqnarray}
$\delta=(\rho_n-\rho_p)/\rho$ is the neutron-proton asymmetry, so the terms corresponding to $\alpha_i$ and $\beta_i$ describe the strong forces in symmetric and asymmetric nuclear matter, respectively. In Refs. \cite{Papakonstantinou:2016zpe,Gil:2019qfr}, it is shown that the optimal numbers of terms are three and four for symmetric and asymmetric parts, respectively, according to which the upper limit of the summation is determined.

In addition to the rules for the expansion of the energy, KIDS formalism also assumes rules for fitting the model parameters. Equation (\ref{eq:kids}) represents the energy in homogeneous infinite nuclear matter. Three $\alpha_i$ and four $\beta_i$ are adjusted to the input nuclear matter data. To apply the theory to finite nuclei, we transform the energy to Skyrme-type contact interactions so that it is easily implemented in
the Hartree-Fock codes. Details for the steps toward finite nuclei are illustrated in Refs. \cite{Gil:2016ryz,Gil:2017tst,Papakonstantinou:2023myk}. In the Skyrme-type force, two terms describing the density gradient and the spin-orbit coupling are added to the nuclear matter functional. Two new parameters in the Skyrme-type force are fitted to reproduce selected nuclear data. Since the fitting to nuclear data is done on top of the parameters determined from the nuclear matter data, nuclear matter EoS can be treated independently of the nuclear properties.

\begin{table*}[tbp]
\begin{center}
\caption{Incompressibility $K_0$ of the symmetric nuclear matter and the symmetry energy parameters $J$, $L$ and $K_{\rm sym}$ (in units of MeV) for the KIDS-A, B, C, D, DD2 and GM1 models.}
\setlength{\tabcolsep}{15.0pt}
\begin{tabular}{ccccccc}
\hline
\hline
 & KIDS-A & KIDS-B & KIDS-C & KIDS-D & DD2 & GM1 \\
\hline
$K_0$ & 230 & 240 & 250 & 260 & 242.7 & 300.5 \\
$J$ & 33 & 32 & 31 & 30 & 31.7 & 32.5 \\
$L$ & 66 & 58 & 58 & 47 & 55 & 94 \\
$K_{\rm sym}$ & $-139.5$ & $-162.1$ & $-91.5$ & $-134.5$ & 93 & 18 \\ 
\hline
\hline
\protect\label{Tab:KIDS}
\end{tabular}
\end{center}
\end{table*}

Following the rules described above, KIDS-A, B, C and D models are determined to satisfy the neutron star properties constrained by modern astronomical observations. The density dependence of the symmetry energy plays a crucial role in determining the nuclear matter EoS at densities below and above the saturation density. With the purpose to reduce the uncertainties in the density dependence of the symmetry energy, incompressibility of the symmetric matter $K_0$, and the symmetry energy parameters $J$, $L$ and $K_{\rm sym}$ are constrained to reproduce the nuclear data and neutron star observation simultaneously. The detailed process is explained in \cite{Gil:2020wct}. Table \ref{Tab:KIDS} tabulates the values of $K_0$, $J$, $L$ and $K_{\rm sym}$ for the KIDS-A, B, C and D models.

\subsubsection{RMF models}
\label{Subsubsec:RMF}

For comparison we also consider two well-known RMF models of two different classes - i) the GM1 \cite{Glendenning:1991es} model with non-linear self couplings and ii) the DD2 \cite{Typel:2009sy} model with density-dependent couplings. The Lagrangian is given as \cite{Xia:2022dvw}
 
\begin{widetext}
\begin{eqnarray} 
\mathcal{L}_{RMF}&=&\bar{\psi}[\gamma_{\mu}(i\partial^{\mu} -g_{\omega}\omega^{\mu} -g_\rho \vec{\rho_\mu}\cdot \vec{\tau}) -(M +g_{\sigma}\sigma)]\psi
+ \frac{1}{2}\partial_{\mu}\sigma\partial^{\mu} -\frac{1}{2}m_{\sigma}^2\sigma^2 -\frac{1}{3}g_2\sigma^3 -\frac{c}{4}g_3\sigma^4  \nonumber\\
&-&\frac{1}{4}\omega_{\mu\nu}\omega^{\mu\nu} +\frac{1}{2}m_{\omega}^2\omega_{\mu}\omega^{\mu} +\frac{1}{4}c_3(\omega_{\mu}\omega^{\mu})^2
-\frac{1}{4} \vec{R}_{\mu \nu} \cdot \vec{R}^{\mu \nu} +\frac{1}{2} m_\rho^2 \vec{\rho_\mu}\cdot \vec{\rho^\mu}.
\label{Eq:rmf_lagrangian} 
\end{eqnarray} 
\end{widetext}
The nucleons interact via the scalar $\sigma$, vector $\omega$ and iso-vector $\rho$ mesons. The vacuum expectation values of the meson fields ($\sigma_0$, $\omega_0$ and $\rho_{03}$) in RMF approximation and the EoS obtained from the Lagrangian (Eq. (\ref{Eq:rmf_lagrangian})) can be found in \cite{Xia:2022dvw}. For the GM1 model the mesons have density-independent couplings $g_{\sigma}$, $g_{\omega}$ and $g_{\rho}$ with the nucleons and $g_2$ and $g_3$ are the higher order scalar field coefficients while $c_3$ is the higher order vector field coefficient. These non-linear self couplings are effectively considered in order to account for the in-medium effects. On the other hand, in the DD2 model \cite{Typel:2009sy} $g_2$=$g_3$=$c_3$=0 and the in-medium effects are treated with the density-dependent couplings $g_i(\rho)$ (where, $i=\sigma, \omega, \rho$) following the Typel-Wolter ansatz \cite{Lu:2011wy}. The values of the density-independent couplings for GM1 model and density-dependent couplings for DD2 model along with the other parameters of the two models can be found in the respective references and also in \cite{Guha:2024pnn}. The saturation properties like the saturation density $\rho_0$, symmetry energy $J$, slope $L$, nuclear incompressibility $K_0$, skewness coefficient $S_0$, and the curvature parameter $K_{\rm{sym}}$ of the nuclear symmetry energy, as obtained for the two RMF models considered in this present work, can be found in the respective references and also in \cite{Xia:2022dvw}. For comparison with the KIDS model, we also display the values of $K_0$, $J$, $L$ and $K_{\rm sym}$ for the DD2 and GM1 models in Table \ref{Tab:KIDS}.

\subsection{Quark models}

\subsubsection{Vector MIT bag (v-Bag) model}
\label{Subsubsec:vBag}

We consider that the quark matter is composed of the $u$, $d$ and $s$ quarks and the electrons. The masses of the quarks are $m_u$ = 5 MeV, $m_d$ = 7 MeV and $m_s$ = 95 MeV. Based on the MIT bag model framework \cite{Chodos:1974je} , in the v-Bag model the repulsion between the quarks is mediated by a vector meson of mass $m_V$ = 783 MeV \cite{Lopes:2021jpm,Kumar:2022byc,Laskos-Patkos:2023tlr,Kumar:2023lhv}. The Lagrangian for the v-Bag model is given by Eq. \ref{Eq:vbag_lagrangian}.
\begin{table*}
\begin{eqnarray}
{\cal L}_{vBag} &=& \sum_{f=u,d,s} \left[\bar{\psi}_f \left\lbrace{\gamma^\mu (i \partial_\mu - g_{qqV} V_\mu) - m_f}\right\rbrace \psi_f - B \right] \Theta (\bar{\psi}_f \psi_f)
\nonumber \\
&+& \frac{1}{2} m^2_V V_\mu V^\mu - \frac{1}{4} V_{\mu \nu}V^{\mu \nu}  + \overline{\psi}_e \big(i \gamma_{\mu} \partial^{\mu} - m_e\big) \psi_e,
\label{Eq:vbag_lagrangian} 
\end{eqnarray}
\end{table*}
$V_\mu$ denotes the vector field, $g_{qqV}$ is the quark-vector meson coupling constant, $B$ is the bag constant, $V_{\mu \nu} = \partial_\mu V_\nu - \partial_\nu V_\mu$ and the Heaviside function $\Theta$=1 inside the bag. For the vector coupling constant, we assume $g_{uuV} = g_{ddV}$ and $X_V=g_{ssV}/g_{uuV}$ = 0.4. In case of QSs, the parameter $G_V=(g_{uuV}/m_V)^2$ and its corresponding value of $B$ are constrained by the Bodmer-Witten conjecture which is related to the stability of the star. For a particular value of $G_V$, $B_{max}$ are determined by the binding energy of the star at the surface $\varepsilon/\rho_B \leq$ 930 MeV while $B_{min}$ is determined with the 2-flavor QM \cite{Farhi:1984qu,Torres:2012xv,Lopes:2021jpm}. The EoS for SQM derived from the Lagrangian (Eq. (\ref{Eq:vbag_lagrangian})) can be found in \cite{Lopes:2021jpm}. The parameter set, used in the present work for obtaining the SQS configurations with the v-Bag model, is tabulated in Tab. \ref{Tab:vbag}.
\begin{table}[!ht]
\caption{Stability window obtained for the vector MIT Bag model with $X_V=$0.4.}
{{
\setlength{\tabcolsep}{15pt}
\begin{center}
\begin{tabular}{ c c c c c c c c}
\hline
\hline
& $G_V$ & $B^{1/4}_{min}$ & $B^{1/4}_{max}$ \\
&(fm$^2$) &(MeV)  &(MeV)  \\
\hline
\hline
v-Bag 1 & 0.3 & 138 & 148 \\
v-Bag 2 & 0.5 & 134 & 143 \\
\hline
\hline
\end{tabular}
\end{center}
}}
\protect\label{Tab:vbag}
\end{table}  
We consider the average value of $B_{max}$ and $B_{min}$ corresponding to each value of $G_V$ in order to obtain the EoS for SQSs as in \cite{Sen:2022pfr}.

\subsubsection{Nambu-Jona-Lasino (NJL) model}
\label{Subsubsec:NJL}

We consider the effective NJL model \cite{Nambu:1961tp} for the 3-flavor SQM. The model includes the scalar, pseudo-scalar, vector, and the t’Hooft six-fermion interaction. The last one is required for the axial symmetry breaking \cite{Hatsuda:1987pc,Hatsuda:1994pi,Kitazawa:2002jop,Lopes:2020rqn,Masuda:2012ed}. The complete Lagrangian is given as
\begin{eqnarray}
{\cal L}_{vNJL} &=& \bar{\psi}_f[\gamma^\mu(i \partial_\mu - m_f)]{\psi}_f \nonumber \\ &+& G_S \sum_{a=0}^8\left[(\bar{\psi}_f \lambda_a \psi)^2 + (\bar{\psi}_f \gamma_5 \lambda_a \psi)^2\right]  \nonumber \\  &-& G_V (\bar{\psi}_f \gamma^\mu \psi)^2 \nonumber \\ &-& K\left\lbrace det[\bar{\psi}(1+\gamma_5)\psi] + det[\bar{\psi}(1-\gamma_5)\psi]\right\rbrace,
\label{Eq:vnjl_lagrangian} 
\end{eqnarray}
where $m_f$ are the current quark masses, $\lambda_a$ are the eight Gell-Mann flavor matrices and $G_S$, $G_V$ and $K$ are the coupling constants. In the NJL model, there is no mediator and the quark-quark interaction is a direct four-Fermi contact interaction. The EoS for SQM derived from the Lagrangian (Eq. (\ref{Eq:vnjl_lagrangian})) can be found in \cite{Lopes:2020rqn,Masuda:2012ed}. We consider the Hatsuda-Kunihiro (HK) parameter set from \cite{Hatsuda:1994pi}, and two values of the vector to scalar coupling ratio ($G_V/G_S$) to describe the SQS configurations.
\begin{table}[!ht]
\caption{Values of the vector to scalar coupling ratio ($G_V/G_S$) for the Nambu-Jona-Lasino quark model.}
{{
\setlength{\tabcolsep}{40pt}
\begin{center}
\begin{tabular}{ c c c c c c c c}
\hline
\hline
& $G_V/G_S$ \\
\hline
\hline
NJL 1 & 0.5 \\
NJL 2 & 1.0 \\
\hline
\hline
\end{tabular}
\end{center}
}}
\protect\label{Tab:NJL}
\end{table}  

\subsubsection{Color-flavor-locked (CFL) model}
\label{Subsubsec:CFL}

In the framework of the MIT bag model \cite{Chodos:1974je}, it is already depicted in \cite{Lugones:2002va} that the ground state of hadronic matter can be strange CFL matter spanning a broad parameter space of the gap parameter $\Delta$ (which is the gap of the QCD Cooper pairs), the bag constant $B$ and the mass of the strange quark $m_s$. To begin with, for the unpaired quark matter (UQM) system, the thermodynamic potential is $\Omega_{free}$ to which the pairing is introduced in terms of $\Delta$ to invoke the CFL phase i.e,
\begin{eqnarray}
\Omega_{CFL}=\Omega_{free} - \frac{3}{\pi^2} \Delta^2 \mu^2 +B,
\label{Eq:CFL_Omega}
\end{eqnarray}
where $\mu$ is the chemical potential. The choice of $\Delta=0$ reduces to the UQM case \cite{Alford:2002rj}. As the equal number of quark flavor contribution is considered, the absence of electrons makes the net matter charge neutral \cite{Rajagopal:2000ff}. In practice, the choice of $B$, $\Delta$ and $m_s$ are not totally independent of each other and is decided by the stability condition of the star. This also suggests that the three flavors of the quarks forming Cooper pairs have same Fermi momentum (and number density) at $T=0$ \cite{Lugones:2002va,VasquezFlores:2017uor}. The EoS of the CFL QM derived from the thermodynamic potential (Eq. (\ref{Eq:CFL_Omega})) can be found in \cite{Lugones:2002va}. The parameter set, used in the present work for obtaining the SQS configurations with the CFL quark model, is tabulated in Tab. \ref{Tab:CFL}.
\begin{table}[!ht]
\caption{Values of gap parameter ($\Delta$) and bag constant ($B$) for the color-flavor-locked quark model.}
{{
\setlength{\tabcolsep}{25pt}
\begin{center}
\begin{tabular}{ c c c c c c c c}
\hline
\hline
& $\Delta$ & $B$ \\
& (MeV)  &(MeVfm$^{-3}$)  \\
\hline
\hline
CFL 1 & 50 & 75 \\
CFL 2 & 100 & 60 \\
\hline
\hline
\end{tabular}
\end{center}
}}
\protect\label{Tab:CFL}
\end{table}  

\section{Hadron-quark phase transition and hybrid star}
\label{Sec:phase transition}

We achieve hadron-quark phase transition and consequently obtain HS configurations with the KIDS-A (softest) and KIDS-D (stiffest) to describe the hadronic phase while the v-Bag model is considered to account for the quark phase. We also considered the other quark models like the NJL and CFL models described in Sec. \ref{Subsubsec:NJL} and \ref{Subsubsec:CFL}, respectively, to obtain HS configurations. However, we found that both NJL and CFL quark models along with the KIDS hadronic model yield unstable HSs. This result is also consistent with \cite{Lopes:2021jpm,Contrera:2016phj,Alaverdyan:2022foz}, where the NJL quark model along with the other hadronic models also gives rise to unstable HS configurations. Similarly, in \cite{Kumar:2023ojk} it can be seen that the CFL quark model combined with the other hadronic models, via Maxwell construction, forms unstable HS configurations. The unstable region of the HSs corresponds to the region where $dM/d\varepsilon_c < 0$, where $\varepsilon_c$ is the central energy density of the star. Therefore, in the present work we consider the configurations of HSs with the KIDS-A, KIDS-D hadronic models and the v-Bag quark model. We present the corresponding results in the context of phase transition and HSs. In case of HSs, it is not mandatory to satisfy the Bodmer-Witten conjecture and the stability conditions in terms of the binding energy per baryon. Also, we showed in \cite{Sen:2024reu} that the values of $G_V$ and $B$ may be constrained in the light of the various astrophysical constraints on the structural properties of the compact stars. Therefore, for obtaining the HS configurations, we fix $X_V$ = 0.4, $B^{1/4}$ = 155 MeV and vary the values of $G_V$.

Assuming that the value of the surface tension at the hadron-quark interface is high ($\geq$ 70 MeV fm$^{-2}$) \cite{Maruyama:2007ss}, phase transition is obtained by employing the Maxwell construction when the pressure $P$ and baryon chemical potential $\mu_B$ of the hadronic and the quark phases become equal i.e,
\begin{eqnarray}
\mu_B^H = \mu_B^Q~~~ ; ~~~ P_H = P_Q.
\label{Eq:Maxwell}
\end{eqnarray}
Therefore, Maxwell construction is characterized by continuous $\mu_B$ while there is a jump in electron chemical potential $\mu_e$ at the interface between the two phases. This leads to a jump in density from the hadronic to the quark phase, pressure being constant within the interval \cite{Contrera:2016phj}. The transition point ($\mu_t,P_t$) of the hadronic and the quark phases in the ($\mu_B - P$) plane corresponds to two specific transition densities - $\rho_t^H$, which signifies the end of the pure hadronic phase and $\rho_t^Q$ that denotes the starting of the pure quark phase in terms of density.

In order to obtain the complete EoS for HNSs and HSs, it is essential to consider the EoS of the crust. For the outer crust, we employ the Baym-Pethick-Sutherland EoS \cite{Baym:1971pw} up to the neutron-drip density and the inner crust is described by the EoS including the pasta phases \cite{Grill:2014aea}.

\section{Structural properties and oscillation modes}
\label{Sec:Structure}

Using the EoS for various composition like HNS matter, SQM, and HS matter, we estimate the global properties such as the gravitational mass ($M$) and the radius ($R$) of the haronic star, SQSs and HSs. The metric for a static spherically symmetric star is given as
\begin{eqnarray}
ds^2=-e^{2\Phi(r)}dt^2 + e^{2\lambda(r)}dr^2 + r^2d\theta^2 + r^2 \sin^2\theta d\phi^2.
\label{Eq:metric}
\end{eqnarray}
Here, $\Phi$ and $\lambda$ are the metric functions. The Einstein field equations are solved for the given metric Eq. (\ref{Eq:metric}) to obtain the Tolman-Oppenheimer-Volkoff (TOV) equations \cite{Tolman:1939jz,Oppenheimer:1939ne} which are given below
\begin{eqnarray}
\frac{dP(r)}{dr}&=&-\Big(\varepsilon(r)+P(r)\Big)\frac{d\Phi(r)}{dr},
\label{Eq:TOV}\\
\frac{d\Phi(r)}{dr}&=&\frac{M(r)+4\pi r^3 P(r)}{r\Big(r-2 M(r)\Big)},
\label{Eq:TOV2}\\
\frac{dM(r)}{dr}&=& 4\pi r^2 \varepsilon(r).
\label{Eq:TOV3}
\end{eqnarray} 
By solving the TOV equations for all possible values of $\varepsilon_c$, we obtain the mass $M$ and the radius $R$ of the star. The mass function $M(r)=r(1-e^{-2\lambda(r)})/2$ is obtained by satisfying Eq. (\ref{Eq:TOV3}). 

We calculate the non-radial oscillations of the hadronic star, SQSs and HSs using the Cowling approximations, which is well formulated in \cite{Sotani:2010mx}. Once we solve the TOV Eqs. (\ref{Eq:TOV} - \ref{Eq:TOV3}), the different oscillation mode frequencies can be obtained by solving two more coupled differential equations along with the corresponding coefficients obtained in terms of the solutions of the TOV equations:
\begin{eqnarray}
\frac{dW(r)}{dr}&=&\frac{d\varepsilon(r)}{dP(r)}\Bigg[\omega^2r^2e^{\lambda(r)-2\Phi(r)}V(r) + \frac{d\Phi(r)}{dr}W(r)\Bigg] \nonumber \\ &-& l(l+1)e^{\lambda(r)}V(r),
\label{Eq:W eqn}\\
\frac{dV(r)}{dr}&=&2\frac{d\Phi(r)}{dr}V(r) - e^{\lambda(r)}\frac{W(r)}{r^2}.
\label{Eq:V eqn}
\end{eqnarray}
The Eqs. (\ref{Eq:W eqn}) and (\ref{Eq:V eqn}) are solved by exploiting the two boundary conditions at the center ($r=0$) and the surface ($r=R$) of the star. At the center ($r=0$) of the star the functions $W(r)$ and $V(r)$ behave asymptotically as
\begin{eqnarray}
W(r)=Ar^{l+1}~{\rm{and}}~~V(r)=-Ar^l/l;
\label{Eq:bc_center}
\end{eqnarray}
where $A$ is an arbitrary constant. At the surface ($r=R$) of the star another boundary condition for $W(r)$ and $V(r)$ is given as
\begin{eqnarray}
\omega^2R^2e^{\lambda(R)-2\Phi(R)}V(R) + \frac{d\Phi(r)}{dr}\Bigg|_{r=R}W(R) = 0.
\label{Eq:bc_surface}
\end{eqnarray}
The Eqs. (\ref{Eq:W eqn}) and (\ref{Eq:V eqn}), are integrated from the center to the surface of the star by assuming an initial value of $\omega^2$. After each integration the value of $\omega^2$ is improved using Ridders' method until Eq. (\ref{Eq:bc_surface}) is satisfied. For both the $f$ and $p$ modes, $l$ = 2. However, the number of nodes is $n$ = 1 for the $p$ mode unlike the $f$ mode which has no node. For HSs, with distinct discontinuity in density, the $g$ mode oscillation frequency is also excited. Thus $g$ mode is present only for HSs and not for HNSs and QSs. In order to account for the density jump for HSs, the following junction conditions are taken into account \cite{Sotani:2010mx} based on the continuous conditions for $W$ and $\Delta P$ :
\begin{eqnarray} 
W_+ = W_-, 
\label{Eq:W junction}
\end{eqnarray}
\begin{eqnarray} 
V_+ = \frac{e^{2\Phi}}{\omega^2 R_g^2} \Bigg\lbrace \frac{\varepsilon_- + P}{\varepsilon_+ + P} \bigg[\omega^2 R_g^2 e^{-2\Phi} V_- + e^{-\Lambda} \frac{d\Phi}{dr} W_- \bigg] \nonumber \\ - e^{-\Lambda} \frac{d\Phi}{dr} W_+ \Bigg\rbrace.
\label{Eq:V junction}
\end{eqnarray}
Here $R_g$ indicates the position of the discontinuity corresponding to the jump in density. $W_-$, $V_-$, and $\varepsilon_-$ are the values of $W$, $V$, and $\varepsilon$ at $r = R_g - 0$ (quark phase) while $W_+$, $V_+$, and $\varepsilon_+$ are the ones at $r = R_g + 0$ (hadronic phase), respectively.

The dimensionless tidal deformability ($\Lambda$) is given as
\begin{eqnarray} 
\Lambda=\frac{2}{3} k_2 R^5.
\label{Eq:Lambda}
\end{eqnarray}
Here $k_2$ is the tidal Love number, which is given in terms of a quantity ($y$) obtained by following \cite{Hinderer:2007mb,Hinderer:2009ca}. For QSs, $y$ is defined 
in terms of the energy density at the surface $\varepsilon_s$ of the QS \cite{Hinderer:2009ca,Kumar:2022byc}. In case of HSs, for the calculation of $y$ and tidal deformability, the jump in density at the hadron-quark interface is taken care of by implementing the correction as suggested by \cite{Takatsy:2020bnx}.
 
\section{Result and discussion}
\label{Sec:Results}

We now present our results and relevant discussions in detail. In the following sections \ref{Subsubsec:Hadronic Neutron Stars}, \ref{Subsubsec:Quark Stars}, and \ref{Subsubsec:Hybrid Stars} we show the numerical outcomes for the HNSs, QSs, and HSs, respectively. In each scenario, before presenting the detailed case studies for the oscillation frequencies, it is important to check whether the considered models are capable of satisfying the constraints on the mass, radius, and tidal deformability of the compact stars. Therefore our narratives are visually comparative in plots while discussing the structural properties of compact stars based on various underlying models in the light of observational findings till date. 

\subsection{Hadronic Neutron Stars}
\label{Subsubsec:Hadronic Neutron Stars}

\begin{figure*}[!ht]
\centering
\subfloat[]{\includegraphics[width=0.49\textwidth]{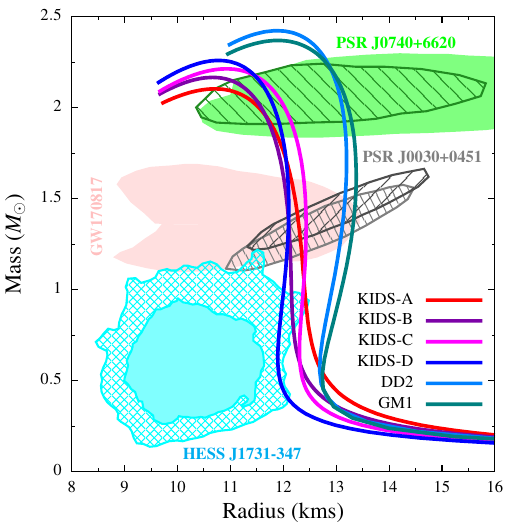}\protect\label{Fig:mr_KIDS}}
\hfill
\subfloat[]{\includegraphics[width=0.49\textwidth]{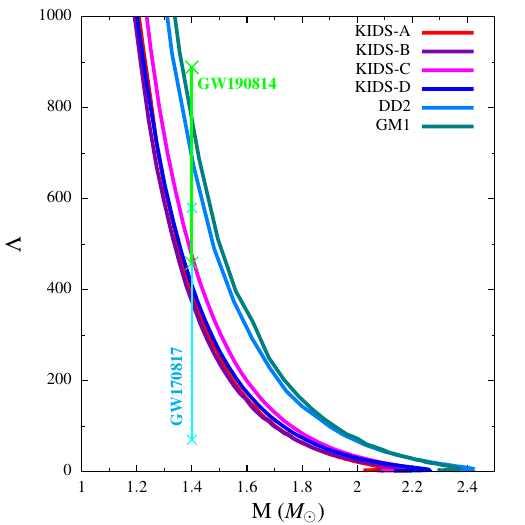}\protect\label{Fig:LamM_KIDS}} \\
\subfloat[]{\includegraphics[width=0.49\textwidth]{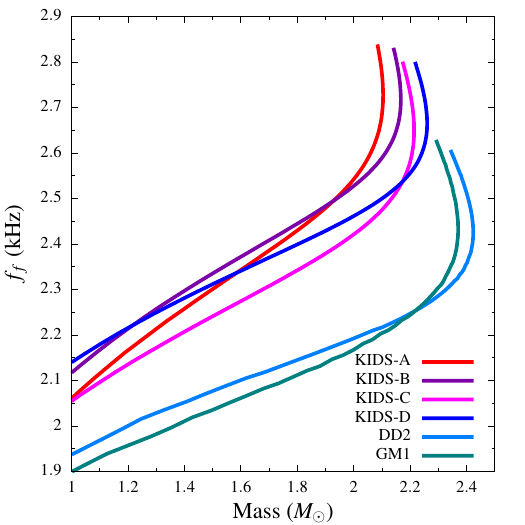}\protect\label{Fig:mf_KIDS}}
\hfill
\subfloat[]{\includegraphics[width=0.49\textwidth]{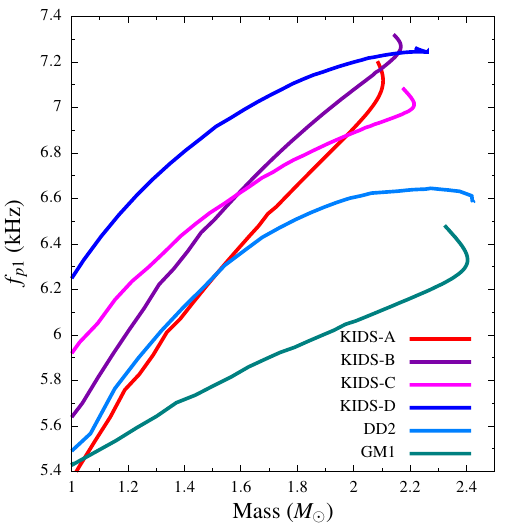}\protect\label{Fig:mp_KIDS}}
\caption{\it (a) Variation of mass with radius of hadronic stars with the KIDS-A, KIDS-B, KIDS-C, and KIDS-D models. Similar variation according to the RMF models DD2 and GM1 are added for comparison. Observational limits imposed from the most massive pulsar PSR J0740+6620 on maximum mass \cite{Fonseca:2021wxt} and corresponding radius \cite{Miller:2021qha,Riley:2021pdl} are also indicated. The constraints on $M-R$ plane prescribed from GW170817 \cite{LIGOScientific:2018cki}, NICER experiment for PSR J0030+0451 \cite{Riley:2019yda,Miller:2019cac} and HESS J1731-347 \cite{Doroshenko:2022} are also compared. (b) Corresponding variation of tidal deformability with mass. The constraint on $\Lambda_{1.4}$ from GW170817 \cite{LIGOScientific:2018cki} and GW190814 \cite{LIGOScientific:2020zkf} are also compared. (c) Corresponding variation of $f$ mode frequency with mass. (d) Corresponding variation of $p_1$ mode frequency with mass.}
\label{Fig:KIDS}
\end{figure*}
\begin{figure*}[!ht]
\centering
\subfloat[]{\includegraphics[width=0.49\textwidth]{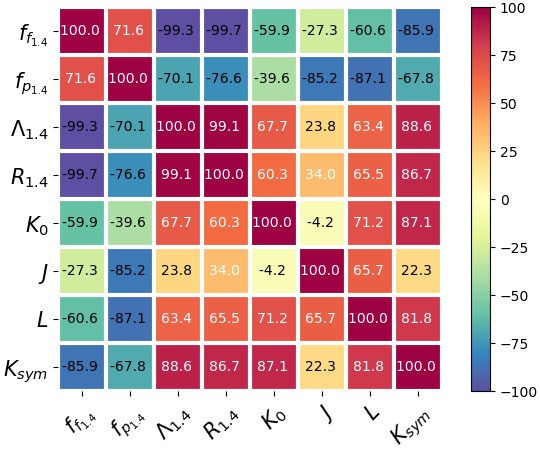}\protect\label{Fig:corr_hadron_1p4}}
\subfloat[]{\includegraphics[width=0.49\textwidth]{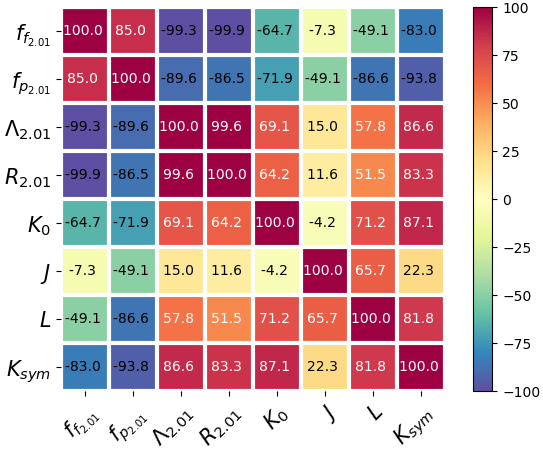}\protect\label{Fig:corr_hadron_2p01}}
\caption{\it Correlation (in percentage) between the nuclear matter parameters, structural, and oscillation properties of hadronic neutron stars of mass (a) 1.4 $M_{\odot}$ and (b) 2.01 $M_{\odot}$.}
\label{Fig:correl_KIDS}
\end{figure*} 

In this section we present the structural and consequently the oscillation properties of the HNSs in Fig. \ref{Fig:KIDS}. The non-relativistic KIDS model yields softer EoS compared to the RMF models. This is reflected in Fig. \ref{Fig:mr_KIDS}, where we find that the maximum mass of the KIDS-A,B,C and D models is less than that of the RMF DD2 and GM1 models. However, the KIDS-A,B,C and D models satisfy the mass and corresponding radius of the most massive pulsar PSR J0740+6620 \cite{Fonseca:2021wxt,Miller:2021qha,Riley:2021pdl}. On the other hand, the radius of the neutron star for a given mass predicted by the KIDS model, is quite less compared to that by both DD2 and GM1 models, making the neutron stars more compact in the framework of the KIDS model. This helps all the four KIDS models satisfy not only the constraints from GW170817 \cite{LIGOScientific:2018cki}, NICER experiment for PSR J0030+0451 \cite{Riley:2019yda,Miller:2019cac} but also HESS J1731-347 \cite{Doroshenko:2022}. The last constraint is not satisfied by the DD2 and GM1 models. In Fig. \ref{Fig:LamM_KIDS} we study the variation of dimensionless tidal deformability with the mass of the HNSs. It can be seen that all the KIDS models A, B, C, and D satisfy the constraint on $\Lambda_{1.4}$ from GW170817 \cite{LIGOScientific:2018cki}. This constraint is not satisfied by the two RMF models considered in this work. Although the constraint on $\Lambda_{1.4}$ from GW190814 \cite{LIGOScientific:2020zkf} is still speculative because the exact nature of the secondary object related to GW190814 emission is still unknown, this constraint is satisfied by the KIDS-C, and the DD2 and GM1 models.

In Figs. \ref{Fig:mf_KIDS} and \ref{Fig:mp_KIDS} we portray the variation of $f$ and $p$ mode oscillation frequencies, respectively, of the HNSs with respect to the mass of the stars. The two RMF models, being stiffer than the KIDS model, have a lower $f$ mode oscillation frequency compared to all the KIDS models. It is also clear from these two figures that the $R_{1.4}$ shows excellent correlation with $f_{f_{1.4}}$ but not with $f_{{p_1}_{1.4}}$. A careful look at Fig. \ref{Fig:mr_KIDS} shows that $R_{1.4}$ for the different hadronic models increases in the trend KIDS-B $<$ KIDS-D $<$ KIDS-A $<$ KIDS-C $<$ DD2 $<$ GM1 while Fig. \ref{Fig:mf_KIDS} shows that $f_{f_{1.4}}$ for the different hadronic models decreases by following the exact trend. Thus, $f_{f_{1.4}}$ follows an inverse correlation with  $R_{1.4}$ as seen in \cite{Kunjipurayil:2022zah}. In the present work the range of $R_{1.4}$ for HNSs is (12.09 - 13.33) km with corresponding range of $f_{f_{1.4}}$ as (2.02 - 2.3). On the other hand, we do not find any well correlation between $R_{1.4}$ and $f_{{p_1}_{1.4}}$ as we see from Fig. \ref{Fig:mp_KIDS} that $f_{{p_1}_{1.4}}$ for the different hadronic models decreases by following the trend KIDS-D $>$ KIDS-C $>$ KIDS-B $>$ DD2 $>$ KIDS-A $>$ GM1. Comparing Figs. \ref{Fig:mf_KIDS} and \ref{Fig:mp_KIDS}, it can be said that the $p$ mode frequency is more affected by the model uncertainties than the $f$ mode. In other words, for any fixed mass of HNS, the difference in $f_f$ is quite less compared to that for $f_{p_1}$ for the different hadronic models. In case of $f$ mode, the difference between $f_{f_{1.4}}^{max}$ (KIDS-B) and $f_{f_{1.4}}^{min}$ (GM1) is 0.28 kHz while for $p_1$ mode the difference is 1.1 kHz. However, it is interesting to note from Fig. \ref{Fig:mp_KIDS} that $f_{{p_1}_{1.4}}$ is well correlated with $L$. $f_{{p_1}_{1.4}}$ follows the exact reverse trend compared to that of $L$ : GM1 $>$ KIDS-A $>$ KIDS-B,C $>$ DD2 $>$ KIDS-D, following Tab. \ref{Tab:KIDS}. Since KIDS-B and C have the same value of $L$, the $f_{{p_1}_{1.4}}$ for these two models are very close, as seen from Fig. \ref{Fig:mp_KIDS}.

We now seek the percentage of linear correlation between the nuclear matter parameters, structural, and oscillation properties of HNSs by calculating the Pearson correlation coefficients. We display the corresponding results in Fig. \ref{Fig:correl_KIDS}. From the results of HNSs of masses 1.4 $M_{\odot}$ and 2.01 $M_{\odot}$ shown in Figs. \ref{Fig:corr_hadron_1p4} and \ref{Fig:corr_hadron_2p01}, respectively, we find that irrespective of the mass of HNSs, both the $f$ and the $p$ mode frequencies are negatively correlated to all the nuclear matter properties like $K_0$, $J$, $L$ and $K_{\rm sym}$. Comparing the results for mass 1.4 $M_{\odot}$ and 2.01 $M_{\odot}$ we notice that among the various nuclear matter properties, $f$ mode frequency is most correlated to $K_{\rm{sym}}$ while $p$ mode frequency shows maximum correlation with $L$, irrespective of mass of HNSs. In both cases, $f$ mode frequency is also moderately correlated to $K_0$ and $L$. $f_p$ and $J$ are moderately correlated, whereas, the correlation between $f_f$ and $J$ is poor. The correlations between the $f$ and $p$ mode frequencies with the other nuclear matter parameters remain inconclusive, as their magnitudes show huge fluctuations with the mass of the HNSs. Considering the structural properties like $\Lambda$ and $R$, we find that $f$ and $p$ mode frequencies show negative correlation with them. $f_f$ shows strongest correlation with both $\Lambda$ and $R$, irrespective of mass of HNSs while for $f_p$ the correlation with $\Lambda$ and $R$ is less compared to that of $f_f$ and also diminishes with decreasing mass of HNSs.

\subsection{Quark Stars}
\label{Subsubsec:Quark Stars}

\begin{figure*}[!ht]
\centering
\subfloat[]{\includegraphics[width=0.49\textwidth]{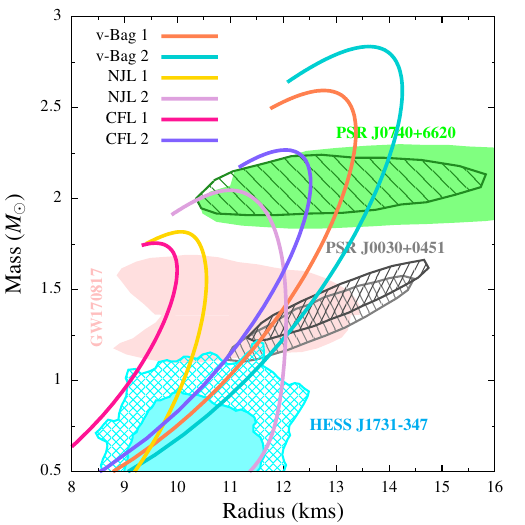}\protect\label{Fig:mr_QS}}
\hfill
\subfloat[]{\includegraphics[width=0.49\textwidth]{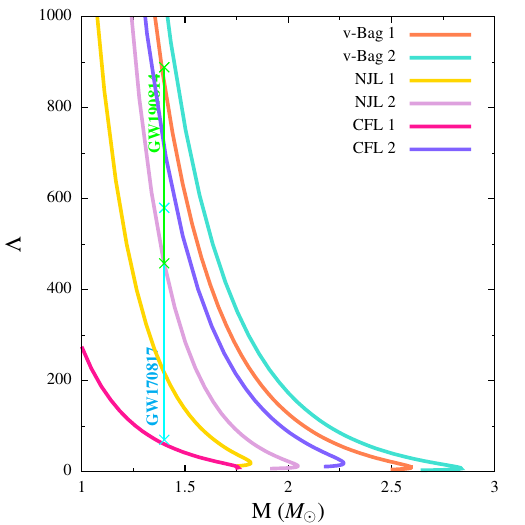}\protect\label{Fig:LamM_QS}} \\
\subfloat[]{\includegraphics[width=0.49\textwidth]{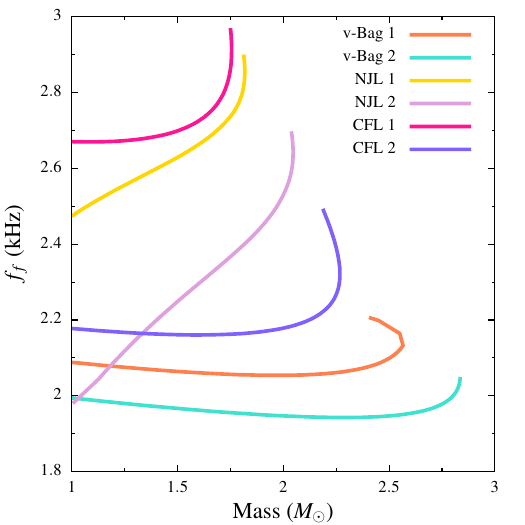}\protect\label{Fig:mf_QS}}
\hfill
\subfloat[]{\includegraphics[width=0.49\textwidth]{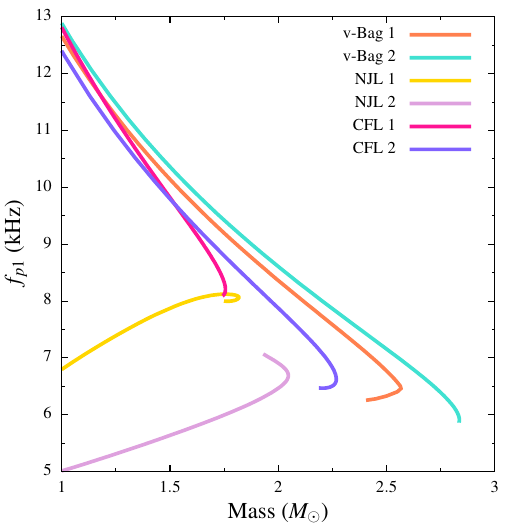}\protect\label{Fig:mp_QS}}
\caption{\it (a) Variation of mass with radius of quark star with the v-Bag, NJL and CFL models. (b) Corresponding variation of tidal deformability with mass. (c) Corresponding variation of $f$ mode frequency with mass. (d) Corresponding variation of $p_1$ mode frequency with mass.}
\label{Fig:QS}
\end{figure*}
\begin{figure*}[!ht]
\centering
\subfloat[]{\includegraphics[width=0.49\textwidth]{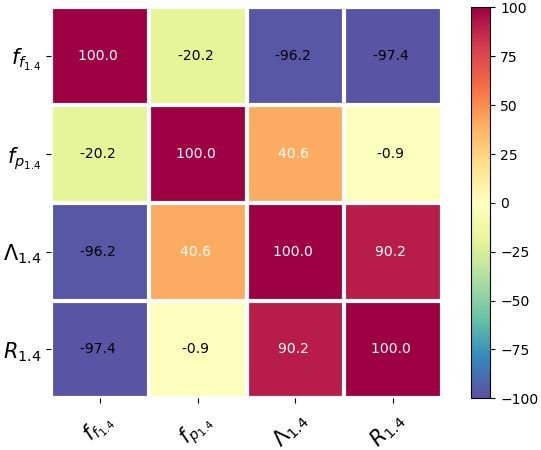}\protect\label{Fig:corr_quark_1p4}}
\subfloat[]{\includegraphics[width=0.49\textwidth]{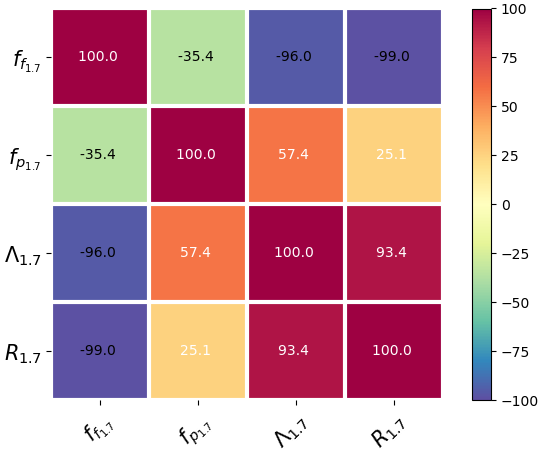}\protect\label{Fig:corr_quark_1p7}}
\caption{\it Correlation (in percentage) between the structural and oscillation properties of quark stars of mass (a) 1.4 $M_{\odot}$ and (b) 1.7 $M_{\odot}$.}
\label{Fig:correl_QS}
\end{figure*} 

Next, we study the structural and oscillation properties of the SQSs in Fig. \ref{Fig:QS}. The mass-radius dependence of the QSs with different models is shown in Fig. \ref{Fig:mr_QS} while corresponding tidal deformability in Fig. \ref{Fig:LamM_QS}. The parameters chosen for the v-Bag, NJL, and CFL models are according to Tabs. \ref{Tab:vbag}, \ref{Tab:NJL}, and \ref{Tab:CFL}, respectively. Our estimates of mass and radius for the v-Bag and NJL models are consistent with \cite{Sen:2022pfr} and \cite{Lopes:2020rqn,Masuda:2012ed}, respectively. In accordance with other works in literature, we find that in the case of v-Bag and NJL models, greater quark repulsion results in more massive SQSs with greater radius while in the case of CFL SQSs higher value of gap parameter yields comparatively massive SQSs with larger radius. Out of all the models, although the v-Bag 1, v-Bag 2, NJL 2, and CFL 2 satisfy all the constraints on the mass-radius plane, the v-Bag 2 satisfies neither the constraint from GW170817 nor from GW190814.

Figs. \ref{Fig:mf_QS} and \ref{Fig:mp_QS} depict the variation of $f$ and $p$ mode non-radial oscillation frequencies of the SQSs, respectively. Comparing the two figures, we find that the $p$ mode is quite sensitive to the low density EoS than the $f$ mode. In absence of a bag constant, the EoS of the NJL model can be obtained from a much lower density compared to those of the v-Bag and CFL quark models. The latter models incorporate the bag constant in the EoS and in the expression of pressure, $B$ appears in a subtractive form \cite{Lopes:2021jpm,Lugones:2002va} that gives an unstable negative pressure range at low density. Thus, we find from Fig. \ref{Fig:mp_QS} that the $p_1$ mode frequency of the NJL model at low mass is quite low compared to that of the v-Bag and CFL models. In other words, the $p$ mode frequency is more dependent on the model considered. Similar to HNS, it can also be seen in the case of QSs that $R_{1.4}$ is well correlated with $f_{f_{1.4}}$ because $R_{1.4}$ increases or $f_{f_{1.4}}$ decreases in the order CFL 1 $<$ NJL 1 $<$ CFL 2 $<$ NJL 2 $<$ v-Bag 1 $<$ v-Bag 2. This correlation is totally lost in case of $f_{{p_1}_{1.4}}$.

In Fig. \ref{Fig:correl_QS} we present the percentage of linear correlation between the structural and oscillation properties of SQSs. The $f$ mode frequency is negatively correlated to both the radius and tidal deformability, irrespective of the mass of SQSs. From the results of SQSs of masses 1.4 $M_{\odot}$ and 1.7 $M_{\odot}$ shown in Figs. \ref{Fig:corr_quark_1p4} and \ref{Fig:corr_quark_1p7}, respectively, we find that $f$ mode frequency of SQSs is very strongly correlated to both radius and tidal deformability, regardless of mass of the SQSs while $p$ mode shows moderate correlation with $\Lambda$ but poor correlation with radius.

\subsection{Hybrid Stars}
\label{Subsubsec:Hybrid Stars}
\begin{figure*}[!ht]
\centering
\subfloat[]{\includegraphics[width=0.49\textwidth]{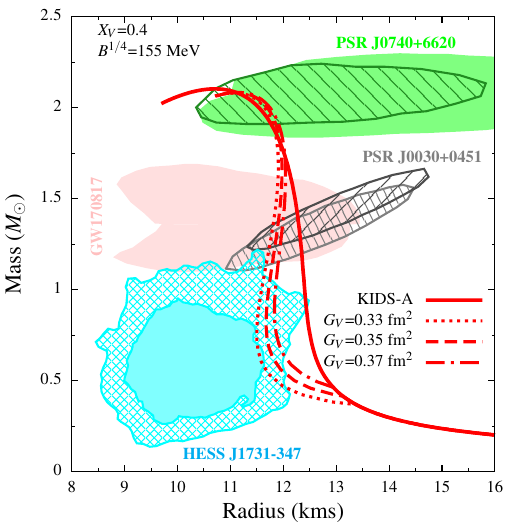}\protect\label{Fig:mr_hybrid_KIDSa}}
\hfill
\subfloat[]{\includegraphics[width=0.49\textwidth]{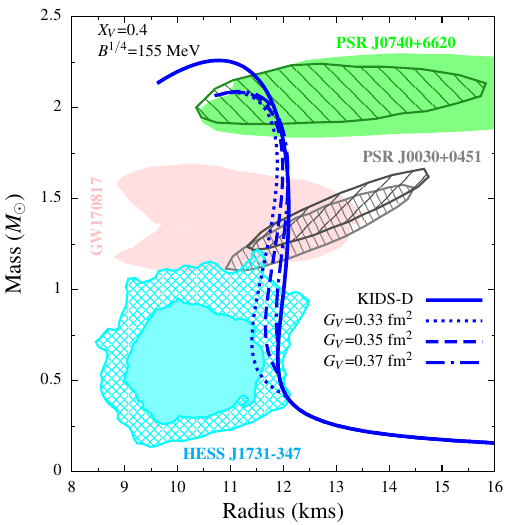}\protect\label{Fig:mr_hybrid_KIDSd}} \\
\subfloat[]{\includegraphics[width=0.49\textwidth]{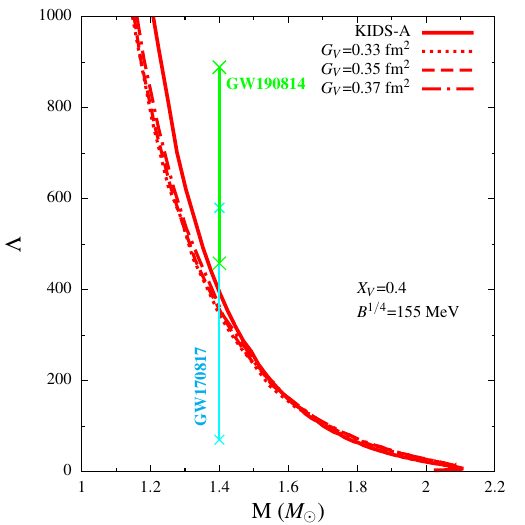}\protect\label{Fig:LamM_hybrid_KIDSa}}
\hfill
\subfloat[]{\includegraphics[width=0.49\textwidth]{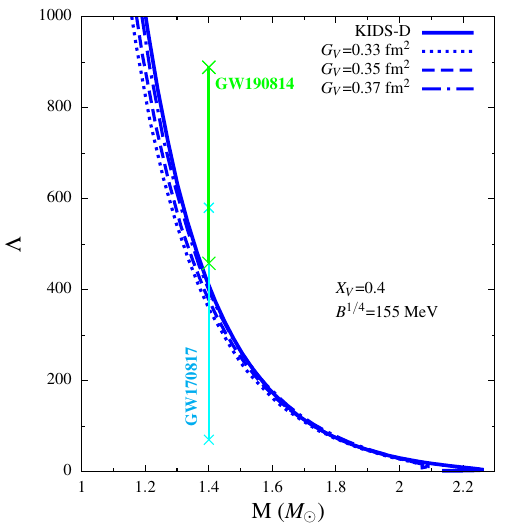}\protect\label{Fig:LamM_hybrid_KIDSd}}
\caption{\it (a) Variation of mass with radius of hybrid star with the KIDS-A and v-Bag models. (b)  Variation of mass with radius of hybrid star with the KIDS-D and v-Bag models. (c) Variation of tidal deformability with mass of hybrid star with the KIDS-A and v-Bag models. (d) Variation of tidal deformability with mass of hybrid star with the KIDS-D and v-Bag models.}
\label{Fig:mr_Lam_hybrid}
\end{figure*}
The structural and oscillation properties of the HSs are next studied. Coherently with \cite{Lopes:2021jpm} it is shown in \cite{Sen:2024reu} that for $X_V$ = 0.4 and $B^{1/4}$ = 155 MeV the phase transition is quite early while considering $G_V<$ 0.4 but the transition point shifts abruptly to a higher value of the chemical potential and pressure (density) at $G_V$ = 0.4 and the HS configurations mostly become unstable for $G_V \geq$ 0.4. In \cite{Sen:2024reu} it is also seen that with the KIDS-A and D hadronic models, and the fixed parameters of the v-Bag model ($X_V$, $B^{1/4}$)=(0.4, 155 MeV), a suitable range for obtaining reasonable HS configurations is 0.3 $\leq$ $G_V$ $\leq$ 0.4 in the light of various astrophysical constraints on the mass, radius, and tidal deformability of the compact stars. Therefore in the present study we chose to work within this range of $G_V$. In Fig. \ref{Fig:mr_Lam_hybrid} we show the results of the structural properties of HSs for the KIDS-A (Figs. \ref{Fig:mr_hybrid_KIDSa} and \ref{Fig:LamM_hybrid_KIDSa}) and KIDS-D models (Figs. \ref{Fig:mr_hybrid_KIDSd} and \ref{Fig:LamM_hybrid_KIDSd}). We observe early phase transition with low transition mass (density). The maximum mass of the HSs is very less affected by $G_V$ and increases slightly with increasing values of $G_V$. However, the values of $R_{1.4}$ and $\Lambda_{1.4}$ increase with increasing $G_V$. Figures \ref{Fig:mr_hybrid_KIDSa} and \ref{Fig:mr_hybrid_KIDSd} show that all the HS configurations satisfy all the different constraints on the mass-radius plane of compact stars. In the $\Lambda$-$M$ plane, all the HS configurations satisfy the constraints from GW170817 but not GW190814.

\begin{figure*}[!ht]
\centering
\subfloat[]{\includegraphics[width=0.33\textwidth]{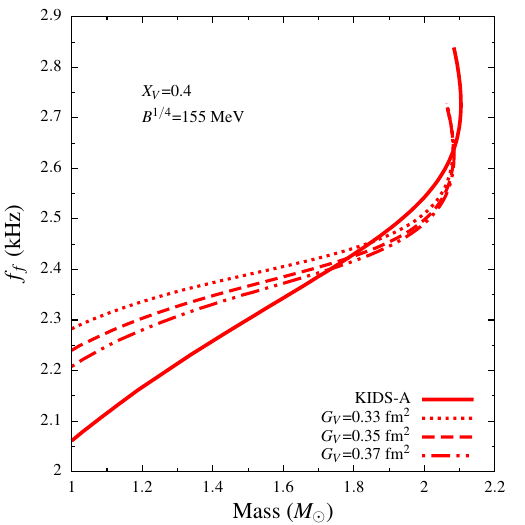}\protect\label{Fig:mf_hybrid_KIDSa}}
\subfloat[]{\includegraphics[width=0.33\textwidth]{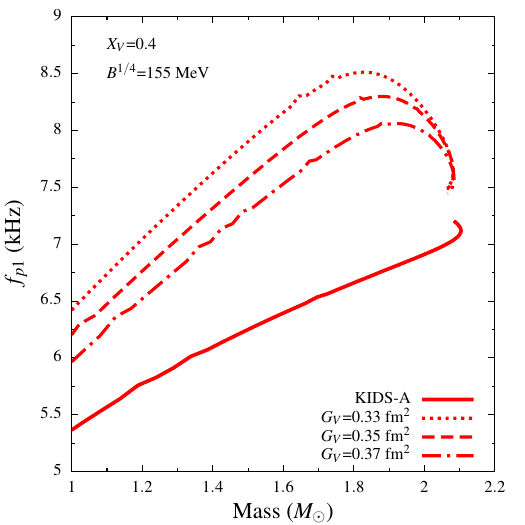}\protect\label{Fig:mp_hybrid_KIDSa}}
\hfill
\subfloat[]{\includegraphics[width=0.32\textwidth]{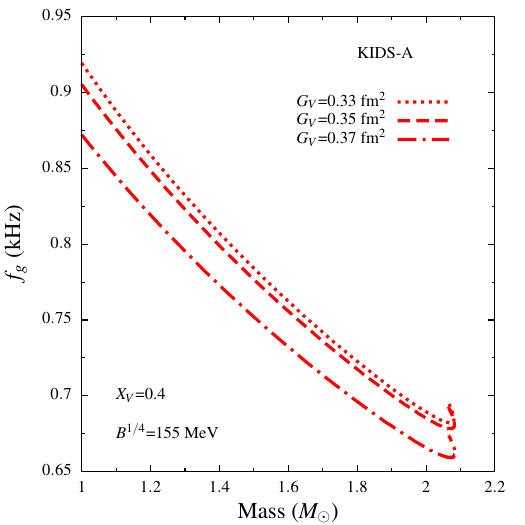}\protect\label{Fig:mg_hybrid_KIDSa}}
\caption{\it Variation of (a) $f$, (b) $p_1$ and (c) $g$ mode frequencies with mass for hybrid stars with KIDS-A model.}
\label{Fig:fpg_KIDSa}
\subfloat[]{\includegraphics[width=0.32\textwidth]{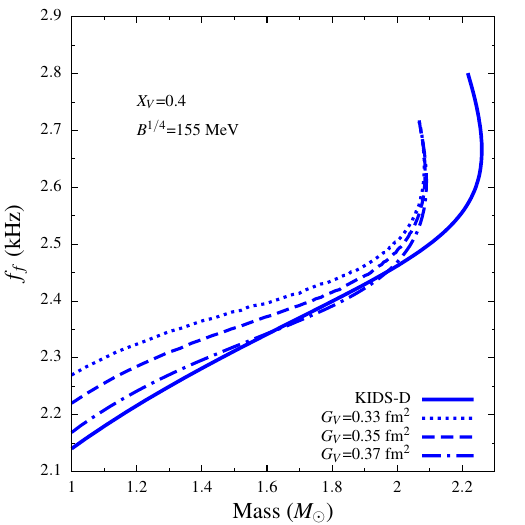}\protect\label{Fig:mf_hybrid_KIDSd}}
\subfloat[]{\includegraphics[width=0.33\textwidth]{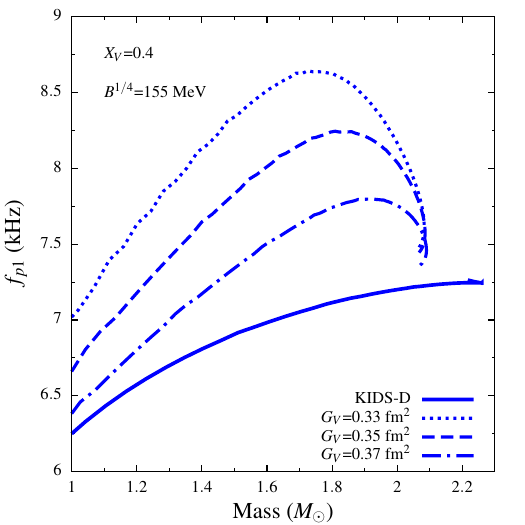}\protect\label{Fig:mp_hybrid_KIDSd}}
\hfill
\subfloat[]{\includegraphics[width=0.32\textwidth]{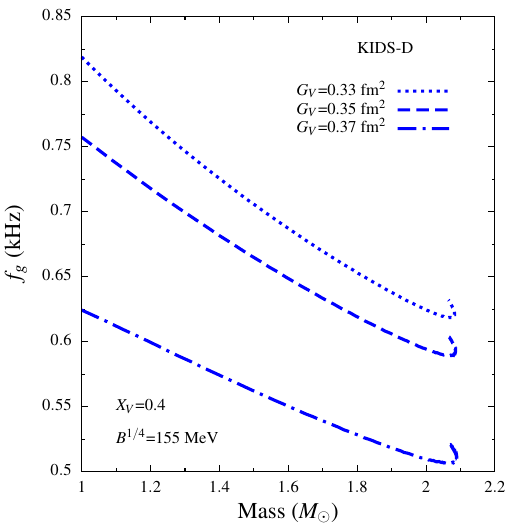}\protect\label{Fig:mg_hybrid_KIDSd}}
\caption{\it Variation of (a) $f$, (b) $p_1$ and (c) $g$ mode frequencies with mass for hybrid stars with KIDS-D model.}
\label{Fig:fpg_KIDSd}
\end{figure*} 

We proceed to display the estimates of $f$, $p_1$, and $g$ mode frequencies of the HSs in the Figs. \ref{Fig:fpg_KIDSa} and \ref{Fig:fpg_KIDSd} for the KIDS-A and KIDS-D hadronic models, respectively. Similar to HNSs and QSs, we observe that the negative correlation between $R_{1.4}$ and $f_{f_{1.4}}$ for HSs in Figs. \ref{Fig:mf_hybrid_KIDSa} and \ref{Fig:mf_hybrid_KIDSd}. Contrary to HNSs and QSs, it is interesting to note that the negative correlation between $R_{1.4}$ and $f_{{p_1}_{1.4}}$ exists for HSs in Figs. \ref{Fig:mp_hybrid_KIDSa} and \ref{Fig:mp_hybrid_KIDSd}. Unlike $f$ mode, the $p_1$ mode frequency of HSs is much greater than that of the HNSs even at the high mass regime. However, due to different uncertainties pertaining to EoS based on different composition of compact star, it is still not possible to comment on whether $p$ mode can distinguish between HNS and QS. For the purpose, the $g$ mode frequency needs to be detected, which can support the possibility of phase transition and the existence of HSs. However, its detection is quite challenging in the recent future due its feeble tidal coupling. Theoretically, $g$ mode appears only in the case of a discontinuity in density indicating a phase transition. In the present work, we obtain well-defined $g$ mode frequencies corresponding to different HS configurations in Figs. \ref{Fig:mg_hybrid_KIDSa} and \ref{Fig:mg_hybrid_KIDSd}. We notice that the $g$ mode frequency decreases slightly with increasing mass and $f_{g_{1.4}}$ is negatively correlated to $R_{1.4}$ and $\Lambda_{1.4}$. Interestingly, there is substantial decrease in $g$ mode frequency with increasing values of $G_V$. Therefore, $g$ mode may also provide an idea of the strength of quark repulsion in HSs. For example, with the KIDS-D model, the change in the values of $f_{f_{1.4}}$, $f_{p_{1.4}}$, and $f_{g_{1.4}}$ for the maximum (0.37) and minimum (0.33) values of $G_V$ are 2.57\%, 11.82\%, and 20.86\%, respectively. Similarly, with the KIDS-A model, the changes in the values of $f_{f_{2.01}}$, $f_{p_{2.01}}$, and $f_{g_{2.01}}$ for the maximum and minimum values of $G_V$ are 0.8\%, 2.32\%, and 3.34\%, respectively. Therefore, irrespective of mass of the HSs, $f_g$ is most affected by the strength of quark repulsion, followed by $f_p$.

\begin{figure*}[!ht]
\centering
{\includegraphics[width=0.5\textwidth]{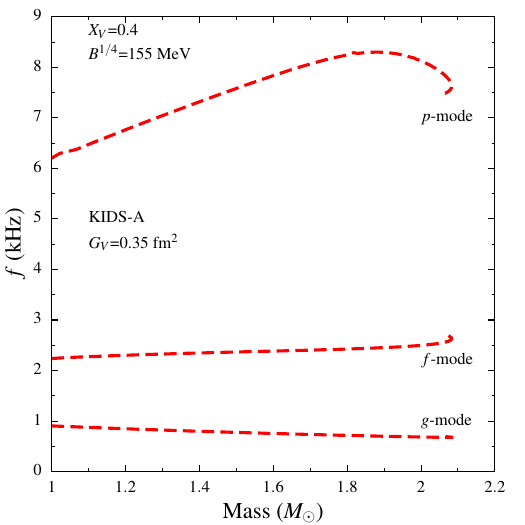}}
\caption{\it Comparison of the range of different modes of frequencies with mass for hybrid stars with KIDS-A model and $G_V$=0.35 fm$^2$.}
\protect\label{Fig:mfpg_hybrid_KIDSa_comparison}\end{figure*} 

In Fig. \ref{Fig:mfpg_hybrid_KIDSa_comparison} we compare the range of $f$, $p$, and $g$ mode frequencies of the HSs for example with the KIDS-A model and $G_V$=0.35 fm$^2$. We find that the magnitudes of the three modes have three distinct ranges. In future the detection of these three modes can be well identified individually. Compared to $f$ and $p$ modes, the $g$ mode frequency is quite feeble and almost independent of mass. 

The hadron-quark interface plays an important role in determining the properties of HSs. Therefore, the properties at the transition are crucial. The estimates of the transition properties such as the transition densities ($\rho_t^H$ and $\rho_t^Q$), transition mass ($M^t$), transition radius ($R^t$) and the $f$, $p_1$ and $g$ mode frequencies at transition ($f_f^t$, $f_{p1}^t$, and $f_g^t$) of HSs are tabulated below in Tab. \ref{Tab:table_trans}.

\begin{table*}[!ht]
\begin{center}
\caption{Transition properties like transition densities, transition mass ($M^t$), transition radius ($R^t$) and the values of $f$, $p_1$ and $g$ mode frequencies at transition ($f_f^t$, $f_{p1}^t$, and $f_g^t$) of hybrid stars with the KIDS-A, D models and v-Bag model with $X_V$ = 0.4 and $B^{1/4}$ = 155 MeV and different values of $G_V$.}
\setlength{\tabcolsep}{10.0pt}
{\small{
\begin{center}
\begin{tabular}{ccccccccc}
\hline
\hline
\multicolumn{1}{c}{Hadronic Model} &
\multicolumn{1}{c}{$G_V$} &
\multicolumn{1}{c}{$\rho_t^H/\rho_0$} &
\multicolumn{1}{c}{$\rho_t^Q/\rho_0$} &
\multicolumn{1}{c}{$M^t$}  & 
\multicolumn{1}{c}{$R^t$}  & 
\multicolumn{1}{c}{$f_f^t$}  & 
\multicolumn{1}{c}{$f_{p1}^t$}  &
\multicolumn{1}{c}{$f_g^t$}  \\
\multicolumn{1}{c}{} &
\multicolumn{1}{c}{(fm$^2$)} &
\multicolumn{1}{c}{} &
\multicolumn{1}{c}{} &
\multicolumn{1}{c}{($M_{\odot}$)}  &  
\multicolumn{1}{c}{(km)}  &  
\multicolumn{1}{c}{(kHz)}  &  
\multicolumn{1}{c}{(kHz)}  & 
\multicolumn{1}{c}{(kHz)}  \\
\hline
KIDS-A &0.33  &1.25  &1.88 &0.38 &13.34 &1.50 &2.63 &1.19 \\
       &0.35  &1.34  &1.91 &0.43 &13.10 &1.58 &3.01 &1.08 \\
       &0.37  &1.44  &1.92 &0.47 &12.92 &1.65 &3.38 &1.02 \\        
\hline
KIDS-D &0.33  &1.50  &1.92 &0.42 &12.05 &1.79 &3.45 &0.97 \\
       &0.35  &1.66  &1.98 &0.54 &11.92 &1.87 &4.24 &0.82 \\
       &0.37  &1.87  &2.06 &0.69 &11.90 &1.98 &5.33 &0.68 \\        
\hline
\hline
\end{tabular}
\end{center}
}}
\protect\label{Tab:table_trans}
\end{center}
\end{table*} 

\begin{figure*}[!ht]
\centering
\subfloat[]{\includegraphics[width=0.49\textwidth]{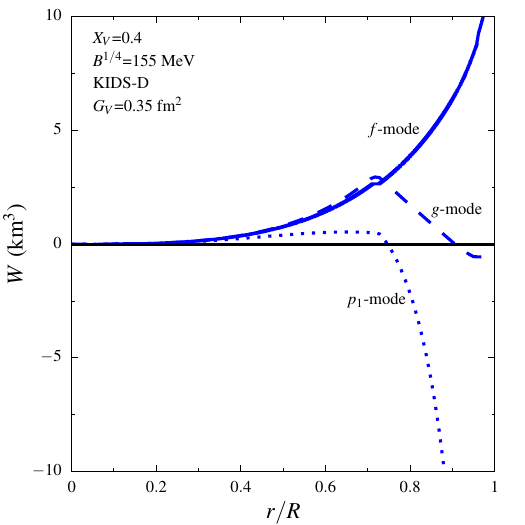}\protect\label{Fig:W_hybrid_KIDSd}}
\subfloat[]{\includegraphics[width=0.49\textwidth]{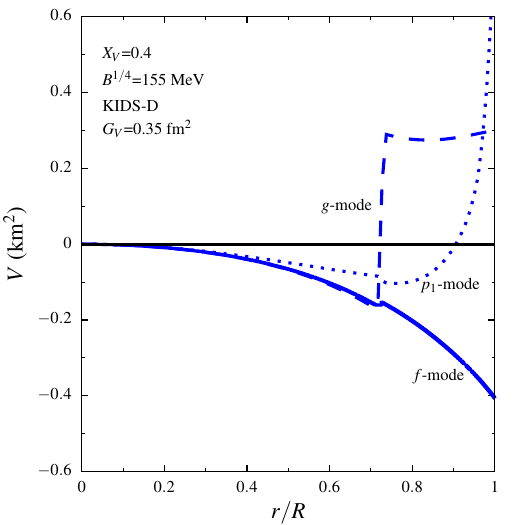}\protect\label{Fig:V_hybrid_KIDSd}}
\caption{\it Radial variation of the eigenfunctions (a) $W$ and (b) $V$ for 1.4 $M_\odot$ hybrid stars with KIDS-D model and $G_V$=0.35 fm$^2$.}
\label{Fig:WV_hybrid_KIDSd}
\end{figure*} 
\begin{figure*}[!ht]
\centering
\subfloat[]{\includegraphics[width=0.49\textwidth]{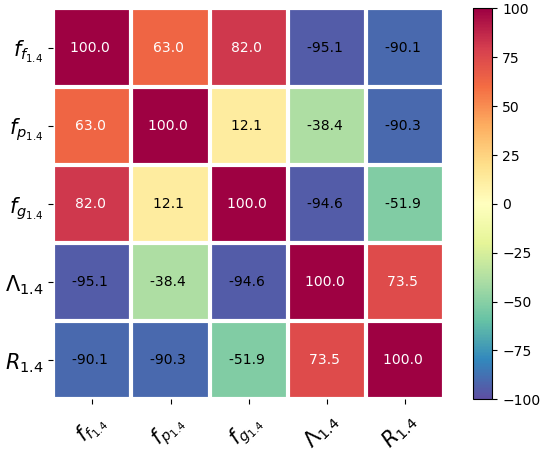}\protect\label{Fig:corr_hybrid_1p4}}
\subfloat[]{\includegraphics[width=0.49\textwidth]{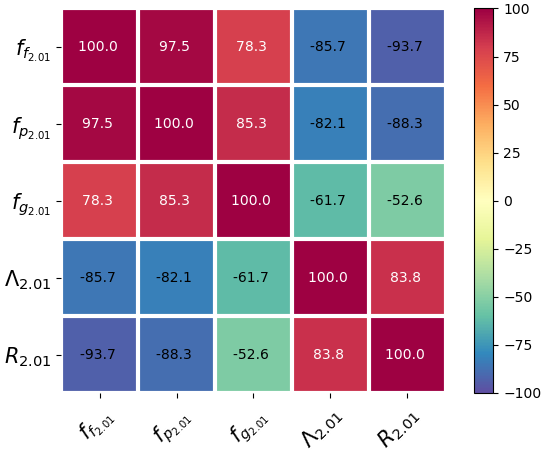}\protect\label{Fig:corr_hybrid_2p01}}
\caption{\it Correlation (in percentage) between the structural and oscillation properties of hybrid stars of mass (a) 1.4 $M_{\odot}$ and (b) 2.01 $M_{\odot}$.}
\label{Fig:correl_HS}
\end{figure*} 

In Fig. \ref{Fig:WV_hybrid_KIDSd} we study the variations of the eigenfunctions $W$ (Fig. \ref{Fig:W_hybrid_KIDSd}) and $V$ (Fig. \ref{Fig:V_hybrid_KIDSd}) with respect to the radius of 1.4 $M_\odot$ HS for example with KIDS-D model and $G_V$=0.35 fm$^2$. The number of nodes corresponding to different modes of oscillation is visibly manifested in the figures. $p_1$ mode contains one node as both the eigenfunctions $W$ and $V$ become zero only once in the range $0< r < R$ for finite $r$, whereas $f$ mode does not have any node and $g$ mode shows phase transition. For $f$ mode, $W$ ($V$) increases (decreases) monotonically from the core to the surface of the star. We observe a glitch around $r/R$ = 0.7 indicating a phase transition. For the $p$ mode, $W$ increases slightly from the core upto $r/R \approx$ 0.7 and then decreases sharply while $V$ decreases from the core slightly up to $r/R \approx$ 0.7 and then shows a stiff increase. For the $g$ mode, the variation of $W$ increases from the core up to $r/R \approx$ 0.7 and then falls drastically whereas $V$ decreases from the core up to $r/R \approx$ 0.7 and then shows a sudden jump and finally a slightly increasing (almost constant) variation towards the surface. This sudden jump in $V$ for the $g$ mode is the most prominent indication of phase transition. This is because at the hadron-quark interface, the junction conditions given by Eqs. (\ref{Eq:W junction}) and (\ref{Eq:V junction}) indicate an abrupt change in $V$ while $W$ remains constant at the point of transition. Therefore, the slightly decreasing part of $V$ from core to $r/R \approx$ 0.7 indicates the pure quark phase, followed by the vertical jump in $V$ indicating the jump in density due to the phase transition and finally the slightly increasing part of $V$ indicates the hadronic phase at higher radius.

The percentage of linear correlation between the structural and oscillation properties of HSs are displayed in Fig. \ref{Fig:correl_HS}. The $f$, $p$, and $g$ mode frequencies are all negatively correlated to both radius and tidal deformability, irrespective of the mass of HSs. From the results of HSs of masses 1.4 $M_{\odot}$ and 2.01 $M_{\odot}$ shown in Figs. \ref{Fig:corr_hybrid_1p4} and \ref{Fig:corr_hybrid_2p01}, respectively, we find that $f$ mode frequency is very strongly correlated to radius, regardless of mass of the HSs. However, the correlation between $f_f$ and $\Lambda$ decreases with increasing mass of the HSs. Like $f_f$, the $p$ mode frequency is also quite strongly correlated to radius, regardless of mass of the HSs. However, $f_p$ shows largely diminishing correlation with $\Lambda$ for decreasing mass. Similar to both $f$ and $p$ mode frequencies, the $f_g$ is also well correlated with radius, regardless of the mass of HSs. However, the correlation between $f_g$ with $\Lambda$ decreases substantially with increasing mass of the HSs.

\begin{figure*}[!ht]
\centering
\subfloat[]{\includegraphics[width=0.49\textwidth]{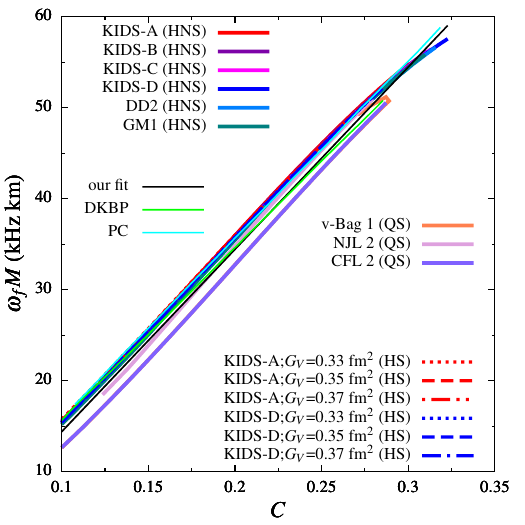}\protect\label{Fig:omgfM_C}}
\subfloat[]{\includegraphics[width=0.49\textwidth]{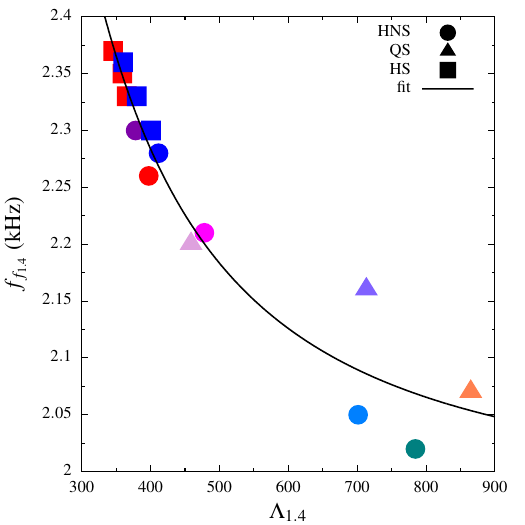}\protect\label{Fig:f1p4_Lam1p4}}
\caption{\it Variation of (a) normalized angular frequency corresponding to $f$ mode with compactness. The fitted results from DKBP \cite{Das:2021dru} and PC \cite{Pradhan:2020amo} are also compared. (b) $f$ mode frequency with respect to tidal deformability of 1.4 $M_{\odot}$ of hadronic, quark and hybrid stars with different models.}
\label{Fig:omgM_C_f1p4_Lam1p4}
\end{figure*} 

Fig. \ref{Fig:omgM_C_f1p4_Lam1p4} is dedicated to the study of universal relations in terms of $f_f$. For this purpose, we consider only those models that satisfy all the astrophysical constraints on the $M-R$ relations in Figs. \ref{Fig:mr_KIDS}, \ref{Fig:mr_QS}, \ref{Fig:mr_hybrid_KIDSa}, and \ref{Fig:mr_hybrid_KIDSd} and constraints on the $\Lambda$-$M$ plane from GW170817 and/or GW190814 in Figs. \ref{Fig:LamM_KIDS}, \ref{Fig:LamM_QS}, \ref{Fig:LamM_hybrid_KIDSa}, and \ref{Fig:LamM_hybrid_KIDSd}. In Fig. \ref{Fig:omgfM_C} we observe that the angular frequency ($\omega_f$) corresponding to the $f$ mode, normalized with the mass, varies universally with the compactness $C$ for the different HNSs, QSs and HSs. This signifies that $f_f$ is almost independent of the composition of the star. Our linear fit for the universal relation is as follows.
\begin{eqnarray}
\omega_f M = aC + b
\end{eqnarray}
where, $a$ = 200.41 and $b$ = -5.65 are the fitting coefficients in the units of kHz-km with dimensionless $C$. Our fit is extremely close to the ones obtained by DKBP \cite{Das:2021dru} and PC \cite{Pradhan:2020amo}, which are obtained for HNSs in the presence of the hyperons using the relativistic Cowling approximation. This emphasizes that the $f$ mode frequency is nearly independent of the composition of the star and the underlying EoS. It is worth mentioning that this characteristic of the $f$ mode frequency is observed even with the calculations involving GR conditions \cite{Zhao:2022tcw}. It is also seen that the mass-scaled $f$ mode angular frequency of QSs is slightly less than that of HNSs and HSs.

In Fig. \ref{Fig:f1p4_Lam1p4} we show that the $f$ mode frequency ($f_{f_{1.4}}$) of the 1.4 $M_{\odot}$ star bears good correlation with $\Lambda_{1.4}$ considering the different compositions and EoSs. 

\begin{figure*}[!ht]
\centering
\subfloat[]{\includegraphics[width=0.49\textwidth]{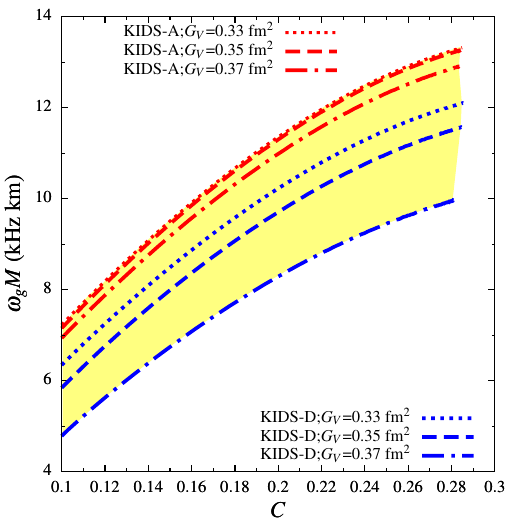}\protect\label{Fig:omggM_C}}
\subfloat[]{\includegraphics[width=0.49\textwidth]{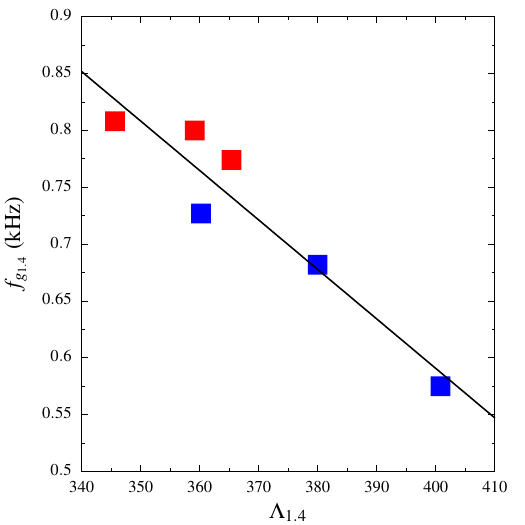}\protect\label{Fig:g1p4_Lam1p4}}
\caption{\it Variation of (a) normalized angular frequency corresponding to $g$ mode with compactness and (b) $g$ mode frequency with respect to tidal deformability of 1.4 $M_{\odot}$ of hybrid stars.}
\protect\label{Fig:omggM_C_g1p4_Lam1p4}
\end{figure*} 

We have already seen that the $f_{{p_1}_{1.4}}$ is not well-correlated with $R_{1.4}$ in case of HNSs and QSs although we find that for HSs $f_{{p_1}_{1.4}}$ is correlated with $R_{1.4}$ and $\Lambda_{1.4}$. Therefore, it is not necessary to test the universality of $\omega_p M$ with respect to $C$ or of $f_{{p_1}_{1.4}}$ with respect to $R_{1.4}$ or $\Lambda_{1.4}$. However, $g$ mode is of special interest in this work. Therefore in Fig. \ref{Fig:omggM_C_g1p4_Lam1p4} we check the universality of $g$ mode frequency. In Fig. \ref{Fig:omggM_C} we show the variation of $\omega_g M$ with respect to $C$ for the HSs. Unlike the universal feature shown by $\omega_f M$ in Fig. \ref{Fig:omgfM_C}, we find $\omega_g M$ spans a much wider region (shaded) that widens more with increasing values of $C$. The result implies that unlike $f$ mode neither $p$ nor $g$ mode shows any universality in terms of the angular frequencies scaled with mass.

In the present work, phase transition is quite early with low transition mass (density), as seen from Tab. \ref{Tab:table_trans}. Therefore, we obtain the canonical stars (1.4 $M_{\odot}$) as hybrid ones. We have already noticed that there exists a negative correlation between $f_{g_{1.4}}$ and $\Lambda_{1.4}$. This is displayed in Fig. \ref{Fig:g1p4_Lam1p4} which also shows that the $f_{g_{1.4}}$-$\Lambda_{1.4}$ relationship (fitted) is also linear.

In this work the oscillation properties are calculated with Cowling approximation. The quantitative results are supposed to deviate from the full GR calculations. However, it is well known that the qualitative results from both the treatments do not vary. This is supported by our correlation results in this work. For example, the correlation results in full GR treatment in \cite{Kunjipurayil:2022zah} also ensure that irrespective of mass of the HNSs, $f_f$ shows strongest correlation with $K_{\rm{sym}}$ while $f_{p_{1.4}}$ has the strongest correlation with $L$ compared to other the nuclear matter parameters. Moreover, the correlation between $f_f$ and $K_0$ is found to decrease slightly with increasing mass of HNSs. In both Cowling approximation and full GR treatment the value is around 60\% in \cite{Kunjipurayil:2022zah}. This value is very close to our estimation of $f_f$-$K_0$ as seen from Fig. \ref{Fig:correl_KIDS}. Similarly, the $f_f$-$L$ correlation decreases with increasing mass in the full GR condition \cite{Kunjipurayil:2022zah}. In the present work, we also find that $L$ is more correlated to $f_{f_{1.4}}$ than $f_{f_{2.01}}$ (Fig. \ref{Fig:correl_KIDS}). Also, it is already seen in \cite{Tonetto:2020bie,Zhao:2022toc} the Cowling approximation is reasonably suitable for the calculation of $g$ mode frequency. 

\section{Summary}
\label{Sec:Conclusion}

In this work, we studied the non-radial oscillation frequencies arising from the quakes in the compact objects such as HNSs, SQSs, and HSs. $f$ and $p$ modes are examined for the HNSs, SQSs, and HSs. $g$ mode frequency is also calculated for the HSs in addition to the $f$ and $p$ modes. The EoS of the HNSs, SQSs, and HSs are strongly dependent on the model. Such a model dependence is thoroughly explored by adopting six models for hadronic matter, three models with varying parameter values for the strange quark matter. For the HSs we consider two hadronic models with three values of repulsion strength of the pure quark matter.

For HNSs, the EoSs of the chosen RMF models are substantially stiffer than the KIDS model. As a consequence, the mass, radius, and tidal deformability of the HNSs are clearly distinguished between the KIDS and the RMF models. The $f$ mode frequency of the 1.4 $M_{\odot}$ star is strongly correlated with $R_{1.4}$. $f_f$ increases exactly in the reverse order of the radius. So we have the variation of $f_{f_{1.4}}$ in the order GM1 $<$ DD2 $<$ KIDS-C $<$ KIDS-A $<$ KIDS-D $<$ KIDS-B. In this work the frequency of the softer hadronic EoS resides in the range (2.2 - 2.3) kHz while the stiffer ones in (2.0 - 2.05) kHz. So, accurate measurement of $f_{f_{1.4}}$ can put a stringent constraint on the stiffness of EoS. The results of $f_{p_{1.4}}$, unlike $f_{f_{1.4}}$, do not show any systematic correlation with the bulk properties of HNSs. However, it is arranged exactly in the reverse order of the slope parameter $L$ of the symmetry energy. In the KIDS model we have $f_{p_{1.4}}$ $\approx$ 6.15, 6.4, and 6.8 kHz with the KIDS-A, C, and D models, respectively. The interval between the models is large enough that if $f_{p_{1.4}}$ is measured accurately, it could provide exclusive constraints to reduce the uncertainty of $L$. 

For the SQSs the bulk properties of the stars depend more strongly on the model. Excluding the models that produce maximum mass below 2 $M_{\odot}$, $f_{f_{1.4}}$ is in the range (2 - 2.2) kHz. So, they are in range similar to that of HNSs. For the $p$ mode, however, only the NJL 2 model gives $f_{p_{1.4}}$ = 5.5 kHz which is similar to that of HNSs, but the other models collectively predict $f_{p_{1.4}}$ in the range (7.5 - 11) kHz. This range is completely separate from that of $f_{f_{1.4}}$. So the measurement of $p$ mode could be useful in identifying the existence of SQSs.

In the calculation of the HSs, in addition to the $f$ and $p$ modes, we consider the $g$ mode that originated from the first order phase transition in the core. The effect of the symmetry energy is accounted for by adopting KIDS-A and KIDS-D models and the uncertainty in the quark matter EoS is incorporated with the strength of the vector repulsion in the v-Bag model. $f_{f_{1.4}}$ of HSs is in the range (2.25 - 2.35) kHz, which overlaps with that of HNSs. So, $f$ mode is not suitable to distinguish between HNSs and HSs. $p$ mode results vary opposite to the $f$ mode. So $f_{p_{1.4}}$ of HSs are clearly distinguished from those of the HNSs, and the dependence of the symmetry and $G_V$ also appears clearly. Thus, the measurement of $p$ mode is expected to provide a diverse and rich story about the various aspects of the theory of dense matter and the properties of the compact objects. $g$ mode shows more dramatic dependence on the symmetry energy. For the KIDS-A model, $f_{g_{1.4}}$ is in the range (0.77 - 0.81) kHz but it is (0.57 - 0.72) kHz for the KIDS-D model. So, there is no overlap between the KIDS-A and KIDS-D models. The dependence on $G_V$ is more prominent in the KIDS-D model. With $G_V$ = 0.33 fm$^2$, $f_{g_{1.4}}$ $\approx$ 0.72 kHz but it drops to 0.57 kHz when $G_V$ = 0.37 fm$^2$. So $g$ mode can provide a unique constraint to reduce the uncertainty in the deconfined quark matter EoS in HSs in terms of quark repulsive strength.

The relation between compactness $C$ and the normalized $f$ mode frequency $\omega_f M$ shows a universal behavior independent of the mass, radius, and composition of the compact objects. Measurement of $f$ mode will thus provide a critical test for the general relativistic description of the compact objects and their properties. We also find strong correlations between the tidal deformability and the $f$ and $g$ mode frequencies of the 1.4 $M_{\odot}$ star, independent of composition. So, they complement each other to sharpen our understanding of the dense matter properties.

Our analysis of the correlation study suggests that disregarding the mass of HNSs $f_f$ is strongly correlated to $K_{\rm{sym}}$, radius, and tidal deformability, moderately correlated to $K_0$ and $L$ and poorly correlated to $J$. $f_p$ shows strongest correlation with $L$ and moderate correlation with radius and tidal deformability. In the case of SQS, irrespective of the mass, $f_f$ is very strongly correlated to both the radius and the tidal deformability, while the $p$ mode frequency for SQSs shows a moderate correlation with $\Lambda$ but a poor correlation with radius. For HSs, regardless of the mass, both $f_f$ and $f_p$ are well correlated to radius. However, only $f_f$ is well correlated to $\Lambda$, regardless of the mass of the HSs. $f_g$ also shows moderate correlation with radius, irrespective of the mass of HSs.

Overall, the non-radial oscillations of the compact objects in the $f$, $p$, and $g$ modes have their own unique and independent characteristics. Measurements of the frequencies are promising to deepen and sharpen our understanding of the strongly interacting infinite matter at densities above the nuclear saturation. From the perspective of the upcoming detectors, the detection of $f_f$ is most plausible while $f_g$ is most challenging. In the present work we have adopted the Cowling approximation since the qualitative results of the $f$ and the $p$ mode frequencies and the quantitative estimation of $f_g$ remain almost the same in full GR treatment. However, it will be interesting to study the oscillation properties of the HNSs, SQSs, and HSs in full GR scenario in a near future work.

\section*{Acknowledgements}

Work of A.G. is supported by the National Research Foundation of Korea (MSIT) (RS-2024-00356960). Work of D.S. is supported by the NRF research Grant (No. 2018R1A5A1025563). Work of C.H.H. is supported by the NRF research Grant (No. 2023R1A2C1003177).

\bibliography{ref}

\end{document}